\documentclass[%
 preprint,
 aip,
rmp,
]{revtex4-2}

\usepackage{graphicx} 
\usepackage{ulem}
\usepackage{float}
\usepackage{hyperref}

\usepackage{xcolor}
\usepackage{amssymb}
\usepackage{dcolumn}

\begin{document}

\title{
Superconductivity in spin-orbit coupled BaBi$_3$ formed by \textit{in situ} reduction of bismuthate films}

\author{Shama}
\author{Jordan T. McCourt}
\author{Merve Baksi}
\author{Gleb Finkelstein}
\author{Divine P. Kumah}
\thanks{Corresponding author. Email: divine.kumah@duke.edu}
\affiliation{%
 Department of Physics, Duke University, Durham, NC 27708, USA
}%

\date{\today}

\begin{abstract} 
Oxygen-scavenging at oxide heterointerfaces has emerged as a powerful route for stabilizing metastable phases that exhibit interesting phenomena, including high-mobility two-dimensional electron gases and high T$_{c}$ superconductivity. We investigate structural and chemical interactions at the heterointerface formed between Al or Eu and the charged-ordered insulator, BaBiO$_3$, leading to emergent superconductivity at 6 K. A combination of X-ray diffraction and electron microscopy measurements shows that oxygen scavenging by the Eu and Al adlayers leads to the formation of superconducting intermetallic BaBi$_3$ in nominal Eu/BaBiO$_3$ and Al/BaBiO$_3$ bilayers. Anisotropic magnetotransport measurements and current-voltage signatures of quasi two-dimensional superconductivity are observed. The mechanisms behind quasi-two-dimensional superconductivity and the role of disorder remain to be clarified. These findings highlight the potential for the use of in situ reduction of bismuthate heterostructures as a platform for stabilizing materials with exotic functional properties. Additionally, the strong spin-orbit coupling at the Bi sites may pave the way for the realization of high T$_{c}$ topological superconductivity.


\end{abstract}

\maketitle

\section{Introduction}
\begin{figure*} 
	\centering
	\includegraphics[width=1\textwidth]{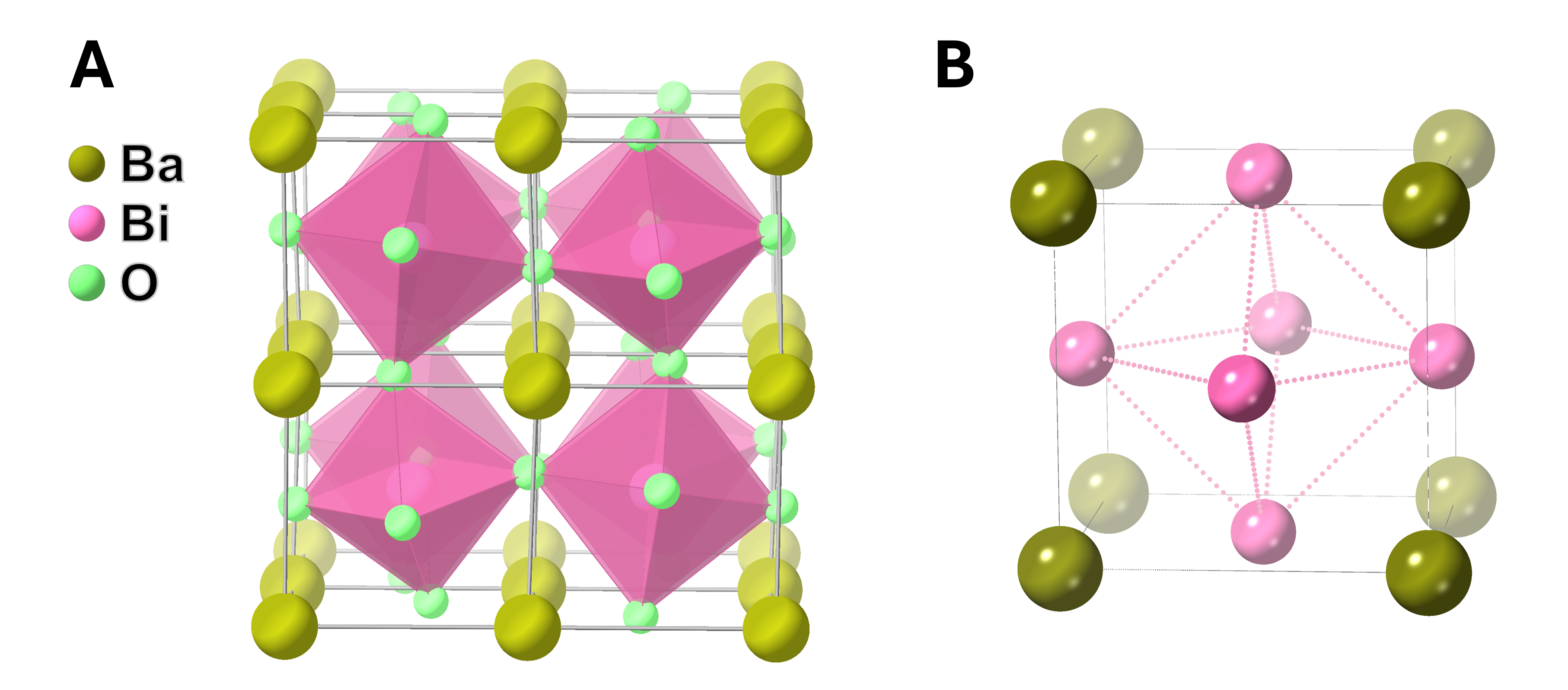} 
	\caption{\textbf{Crystal structure of BaBiO$_3$ and BaBi$_3$}
		 \textbf{(A)} Crystal structure of charged ordered insulating monoclinc BaBiO$_3$ \textbf{(B)} Crystal structure of superconducting tetragonal BaBi$_3$.}
	\label{fig:atomicstructure} 
\end{figure*}

The pursuit of new materials exhibiting high-temperature superconductivity remains a vibrant and active area of research. Following the discovery of high T$_{c}$ superconductivity in cuprates, attention expanded towards the other oxide systems, including bismuthates\cite{sleight1975high, mattheiss1988superconductivity, cava1988superconductivity, hinks1988synthesis}. Notably, superconductivity was also identified in potassium (K) and lead (Pb) doped BaBiO$_{3}$ with a superconducting transition temperature (T$_{c}$) reaching up to 32 K for Ba$_x$K$_y$BiO$_3$ \cite{cava1988superconductivity,mattheiss1988superconductivity,pei1990structural}. In contrast to high-T$_{c}$ cuprates, where conduction is confined to two-dimensional CuO$_{2}$ planes, the BiO$_{2}$ planes in BaBiO$_{3}$ form a three-dimensional network of corner-sharing octahedra, as shown in Figure \ref{fig:atomicstructure}A. The parent compound, monoclinic perovskite-BaBiO$_{3}$ (p-BBO), is known to be a charge density wave insulator with an associated structural breathing mode distortion of the oxygen octahedral cage \cite{cox1976crystal, sleight2015bismuthates, foyevtsova2015hybridization}. Upon doping with K or Pb, this breathing mode is suppressed, leading to the emergence of metallicity and superconductivity \cite{mattheiss1988superconductivity, cava1988superconductivity, pei1990structural, sleight2015bismuthates}.

Interest in Bi-rich superconductors continues due to the potential role of the strong spin-orbit coupling of Bi-6p electrons in enhancing superconductivity, and magnetic field-induced transitions from conventional s-wave to unconventional nodal d-wave superconductivity \cite{shao2016spin, powell2025multiphase}. Intermetallic \textit{A}Bi$_3$ compounds (where \textit{A} = Ba or Sr) have received attention because they exhibit superconductivity enhanced by spin-orbit coupling\cite{matthias1952search, haldolaarachchige2014superconducting, shao2016spin, jha2016hydrostatic}. Theoretical studies show that spin-orbit coupling suppresses strong Fermi nesting and increases electron-phonon coupling matrix elements associated with superconductivity\cite{shao2016spin}. At ambient pressure, BaBi$_3$ crystallizes in a tetragonal structure (\textit{P4/mmm}, a=5.06 $\mathring{A}$, c=5.13 $\mathring{A}$)\cite{jha2016hydrostatic, haldolaarachchige2014superconducting, wang2021superconducting, haberkorn2023superconducting}, featuring corner-sharing Bi$_6$ octahedra (Figure \ref{fig:atomicstructure}B). Magnetotransport measurements on single crystals of BaBi$_3$ and SrBi$_3$ reveal superconducting transitions with onset temperatures of approximately 5.9 K and 5.6 K, respectively \cite{haldolaarachchige2014superconducting, jha2016hydrostatic}. Furthermore, scanning tunneling microscopy (STM) studies demonstrate the effect of symmetry breaking and disorder on the superconducting state in BaBi$_3$ epitaxial films, grown in situ by molecular beam epitaxy\cite{wang2018surface}. However, to date, there have been no reports on the magnetotransport properties of thin BaBi$_3$, due to its high air sensitivity and the availability of lattice matched substrates\cite{jha2016hydrostatic}.

Here, we report the emergence of superconductivity at approximately 6 K in nominally Eu-capped p-BBO thin films grown via molecular beam epitaxy. X-ray diffraction measurements reveal that oxygen scavenging by the in-vacuo deposited metallic Eu or Al overlayer leads to the reduction of the monoclinic p-BBO layers and the decomposition of the reduced BBO films into tetragonal superconducting BaBi$_3$ and rock-salt BaO. Temperature-dependent transport measurements, including angle-dependent magnetotransport and current-voltage characteristics, confirm the quasi-two-dimensional nature of the superconductivity. These findings highlight the potential of utilizing \textit{in situ} reduction of bismuthate heterostructures as a platform for exploring novel quantum states. The strong spin-orbit coupling at Bi sites, may pave the way for the realization of high-temperature topological superconductivity, a long-sought phenomenon in perovskite-based systems\cite{zhang2022tailored, li2015topological, yan2013large}.

\section*{Results and Discussion}

\subsection*{Synthesis}
\begin{figure*} 
	\centering
	\includegraphics[width=1\textwidth]{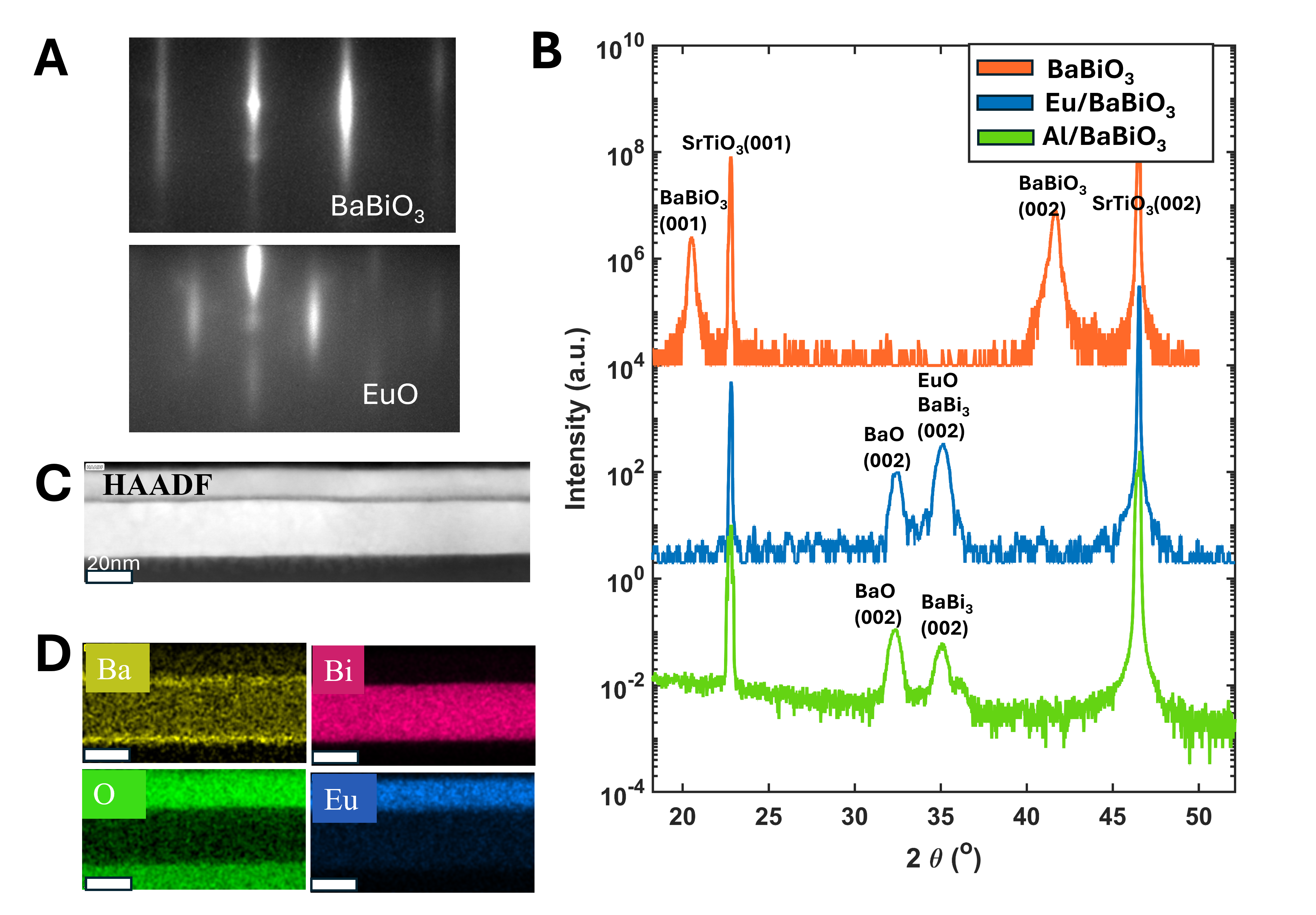} 
	\caption{\textbf{Structural properties of nominal Eu/BaBiO$_3$ heterostructures grown by molecular beam epitaxy.}
		\textbf{(A)} In situ reflection high-energy electron diffraction (RHEED) images measured during the growth of BaBiO$_3$ and Eu metal along the SrTiO$_3$ [100] zone axis. \textbf{(B)} Comparison of X-ray diffraction measurements for a BaBiO$_3$ film on (001)-oriented SrTiO$_3$, a nominal Eu/BaBiO$_3$, and Al/BaBiO$_3$ heterostructures on SrTiO$_3$. \textbf{(C)} Scanning transmission electron microscope high-angular annular dark field (HAADF) image of cross-section of nominal Eu/BaBiO$_3$ heterostructure on SrTiO$_3$. \textbf{(D)} Energy-dispersive X-ray spectroscopy maps for cross-section in (C) showing elemental distributions of Ba, Bi, O, and Eu.}
	\label{fig:structure} 
\end{figure*}

The samples were fabricated in a custom-oxide molecular beam epitaxy system. Bi, Ba, Eu, and Al were evaporated from effusion cells, with their growth rates calibrated by a quartz crystal monitor (QCM). p-BBO thin films of thicknesses ranging from 13 nm (30 unit cells) to 36 nm (90 unit cells) were grown on (001)-oriented SrTiO$_3$ (STO) substrates at a substrate temperature of 585$^{\circ}$ C in an oxygen plasma with an oxygen partial pressure of 6.5x10$^{-6}$ Torr and a plasma power of 250 W. Sharp reflection high-energy electron diffraction (RHEED) diffraction streaks observed during p-BBO growth shown in Figure \ref{fig:structure}A, are indicative of atomically flat p-BBO layers. The in-plane lattice parameter determined from the RHEED spacings is 4.35 $\mathring{A}$, in agreement with bulk stoichiometric p-BBO \cite{pei1990structural}. Due to the large 11\% lattice mismatch between p-BBO and STO, p-BBO grows relaxed via domain matching epitaxy.\cite{zapf2018domain} For the nominal Eu/p-BBO samples, after growth, the p-BBO films were cooled to 500$^{\circ}$ C for the deposition of Eu metal in vacuo. RHEED patterns measured during Eu growth (Figure \ref{fig:structure}A) are consistent with the formation of cubic EuO from the topotactic reduction of the p-BBO layer \cite{qiao2024observation, guo2018euo, mairoser2015high}. The films were cooled in-vacuo and capped with 3 nm AlO$_x$ at room temperature to minimize oxidation of the EuO when exposed to ambient conditions\cite{averyanov2015direct}. A second set of samples was fabricated without Eu. Al was deposited at 500 $^\circ$C in vacuum, after the depostion of p-BBO, and the samples were subsequently cooled to room temperature to form nominal Al/p-BBO bilayers.

\subsection*{X-ray Diffraction (XRD) Measurements}
Figure \ref{fig:structure}B shows a comparison of specular scans along the (00L) direction for the nominal Eu/p-BBO sample and a single-component p-BBO film on STO. For the single-component p-BBO film, Bragg peaks are observed for the STO substrate and stoichiometric p-BBO. The calculated out-of-plane lattice constant for the p-BBO film is determined to be 4.36 $\mathring{A}$ in agreement with the measured pseudocubic values for the bulk monoclinic  phase \cite{pei1990structural}. In contrast to the single-component p-BBO film, no p-BBO Bragg peaks were observed for the nominal Eu/p-BBO sample. Bragg peaks are observed at $ 2\theta=$ 32.316$^o$ and 34.997$^o$ corresponding to the (002) Bragg peaks of rock-salt BaO and face-centered tetragonal BaBi$_3$, respectively. The calculated out-of-plane lattice constants are 5.52 $\mathring{A}$ and 5.11 $\mathring{A}$ in good agreement with the bulk lattice constants for BaO (5.53 $\mathring{A}$) and BaBi$_3$ (5.13 $\mathring{A}$) \cite{zollweg1955x,haldolaarachchige2014superconducting}. The lattice constant of NaCl-type EuO is 5.1439 $\mathring{A}$,\cite{matthias1961ferromagnetic} hence, the EuO (002) Bragg peak will overlap with the BaBi$_3$ (002) peak.

To confirm the formation of BaBi$_3$, XRD measurements were carried out on samples with Al metal deposited directly on a p-BBO film at 500$^\circ$C. We find that the Al metal effectively reduces the p-BBO film to form BaBi$_3$ and BaO as evidenced by XRD measurements in Figure \ref{fig:structure}B. Reciprocal space maps (RSMs) around the STO (102), BaO (113), and BaBi$_3$ (113) Bragg peaks are shown in Figure S1 in the Supplementary Materials\cite{supplement}.  The BaBi$_3$ is oriented with the BaBi$_3$ [0 0 1] axis parallel to the SrTiO$_3$ [0 0 1] axis and the BaBi$_3$ [1 1 0]//SrTiO$_3$ [1 0 0] axis. The lattice constants are determined to be a= 5.21 $\mathring{A}$ and c=5.15 $\mathring{A}$ in agreement with bulk values of $\alpha$-BaBi$_3$ \cite{wang2021superconducting}. The RSM confirms that the BaBi$_3$ is fully relaxed while the BaO is lattice-matched to the STO substrate. The lattice relaxation of BaBi$_3$ is expected due to the 8\% mismatch between the (110) planes with $d_{110}=3.68 \> \mathring{A}$  and STO ($a=3.905\> \mathring{A})$.

Next, we investigate the effect of the ratio of the thickness of the Al and BaBiO$_3$ on the formation of the BaBi$_3$ phase. We compare the X-ray diffraction patterns of 45 monolayers (ML) (1 ML p-BBO = 4.35 $\mathring{A}$) of BaBiO$_3$ capped with 0, 10, and 45 ML of Al in Figure S2 in the Supplementary Materials \cite{supplement}. For the 10 ML Al/45 ML BaBiO$_3$ sample, the perovskite BaBiO$_3$ shifts to lower angles consistent with the formation of partially reduced BaBiO$_{3-x}$ and an increase of the c-axis lattice parameter \cite{cao2021realization}. The perovskite phase is suppressed in the 10 ML Al/45 ML BaBiO$_3$ sample, confirming that further reduction of the perovskite BaBiO$_3$ phase leads to decomposition in BaO and BaBi$_3$.

\subsection*{Scanning transmission electron microscopy analysis}
Scanning transmission electron microscopy (STEM) measurements were performed on the films to further investigate the structural properties of the heterostructures. Figure \ref{fig:structure}C shows a high-angle annular dark field (HAADF) STEM image of a cross-section of a nominal Eu/p-BBO film on STO. Sharp interfaces are observed between the nominal layers, indicative of insignificant intermixing. The energy-dispersive X-ray spectroscopy (EDX) profiles of the section are shown in Figure \ref{fig:structure}D. The EDX maps reveal significant chemical segregation in the nominal p-BBO region. The nominal Eu and STO regions are oxygen-rich, while the nominal p-BBO section is oxygen-poor, consistent with the reduction of p-BBO by the Eu adlayer. Ba-rich regions are observed at the nominal p-BBO/STO and Eu/p-BBO interfaces. In the nominal p-BBO region, we observe lateral phase-separated Ba and Bi-rich regions, as shown in Figure S3 in the Supplementary Materials\cite{supplement}. The oxygen content in the Ba-rich regions is significantly higher than in the Bi-rich sections, consistent with the BaO and BaBi$_3$ Bragg peaks observed in the XRD measurements.

\subsection*{Transport Measurements }

\begin{figure*} 
	\centering
	\includegraphics[width=1\textwidth]{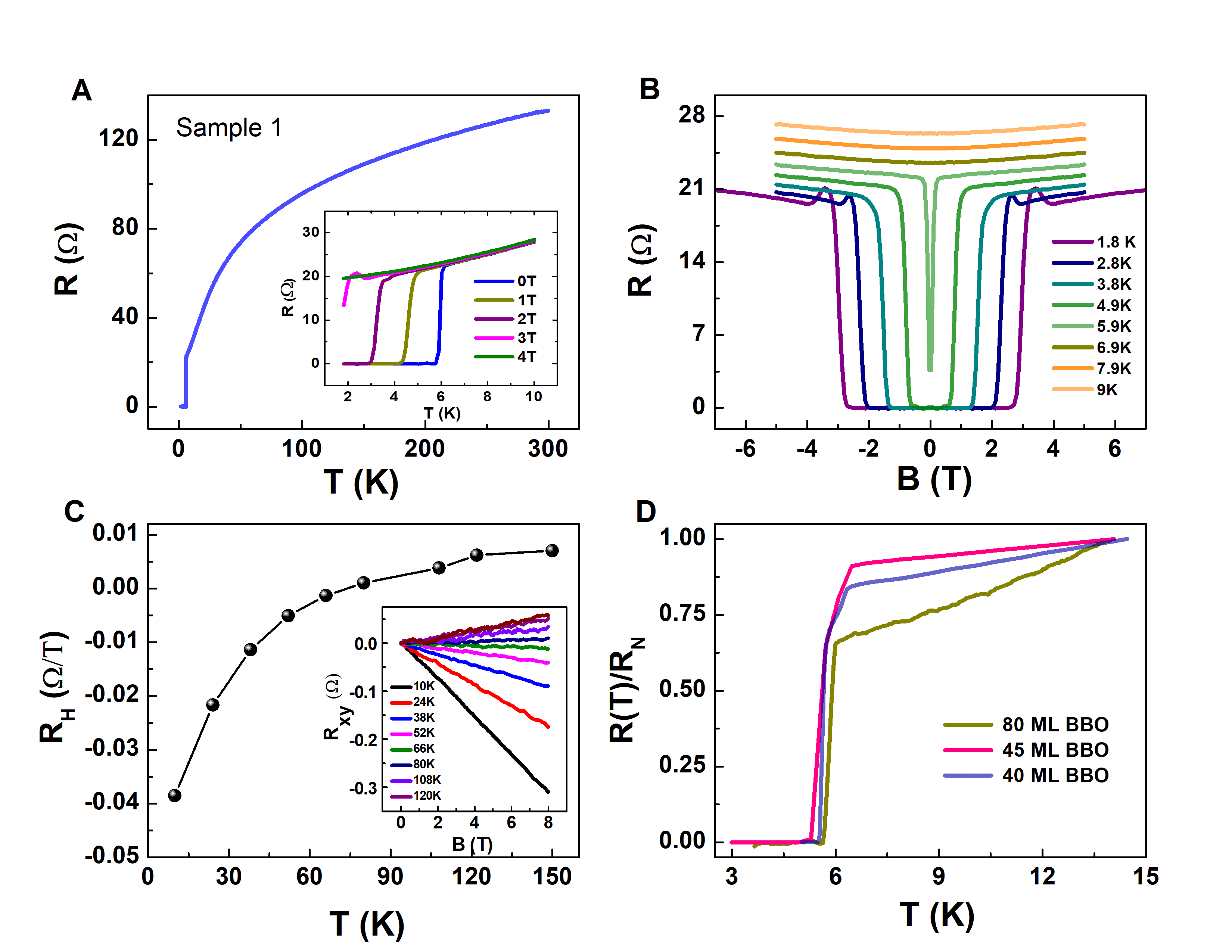} 
		\caption{\textbf{Transport properties of nominal Eu/BBO heterostructures (Sample 1) grown by molecular beam epitaxy.}
		 \textbf{(A)} Resistance versus temperature. The inset graph shows the low temperature resistance as a function of magnetic fields applied perpendicular to the sample surface. \textbf{(B)} Magnetoresistance measured as a function of temperature above and below the superconducting transition. \textbf{(C)} Hall coefficient as a function of temperature above the superconducting transition temperature (T$_c$). The inset shows the Hall resistance vs magnetic field. \textbf{(D)} Thickness dependence of superconducting transition.}
	\label{fig:transport1} 
\end{figure*}

Next, we investigate the transport properties of nominal Eu/p-BBO heterostructures from 1.8 - 300 K. Figure \ref{fig:transport1}A shows the temperature-dependent resistance of a nominal Eu (16 nm)/p-BBO (36 nm) heterostructure (Sample 1). As the temperature decreases to 7 K, the resistance decreases, exhibiting characteristic metallic behavior. The negative curvature of the resistivity between 50-80 K is consistent with the transport properties of bulk BaBi$_3$ and may be associated with two-band conductivity \cite{haldolaarachchige2014superconducting, wang2018two, zverev2009transport}. Between 25 K and 6 K,  the resistance exhibits a linear temperature dependence. This linearity is more clearly demonstrated by the linear fitting of the resistance vs. temperature data shown in Figure S4 (A). This behavior is indicative of strange-metal characteristics commonly observed in unconventional superconductors \cite{Yuan2022strange}.
Upon further cooling, the sample undergoes a transition to a superconducting state with a narrow transition width of approximately 0.133 K, reaching a zero resistance state at 5.77 K ($T_{c,0}$). The onset superconducting temperature $T_{c,onset}$ (defined as the temperature at which the resistance drops to 90\% of its normal state value), is 6.03 K, significantly higher than the previously reported value of 3.5 K for EuO/BaBiO$_3$ heterostructures\cite{qiao2024observation} but in agreement with the reported values for bulk BaBi$_{3}$\cite{haldolaarachchige2014superconducting, jha2016hydrostatic, shao2016spin, wang2021superconducting, haberkorn2023superconducting}. Above T${_c}$, for B = 4T, linearity is suppressed along with superconductivity, consistent with previous reports \cite{haldolaarachchige2014superconducting,saito2016superconductivity,chen2021superconductivity}. We performed a T$^{2}$ fit, characteristic of Fermi-liquid behavior, to the resistivity curve for B = 4 T as shown in Figure S4 (B). A clear deviation from T$^{2}$ behavior is observed above 10K, in agreement with transport measurements for bulk BaBi$_{3}$ \cite{haldolaarachchige2014superconducting}. These results combined with XRD results suggest the emergence of superconductivity is a result of the formation of BaBi$_3$.

The inset of Figure \ref{fig:transport1}A shows the temperature-dependent resistance of Sample 1 with a magnetic field ($B_\perp$), applied out of the plane. $T_{c,onset}$ is suppressed as $B_\perp$ increases, further confirming superconductivity. Magnetoresistance curves with $B_\perp$ at different temperatures close to the superconducting transition are plotted in Figure \ref{fig:transport1}B. In the normal state above T$_{c}$, the resistance exhibits a quadratic dependence on the applied magnetic field, similar to conventional metals\cite{abrikosov2017fundamentals}. Below T$_{c}$, a resistance peak appears above the critical magnetic field, suggesting disorder arising from granularity\cite{jaeger1989onset}. As the BaBi$_3$ phase emerges from the decomposition of p-BBO into BaO and BaBi$_3$, inhomogeneity and defect phases are likely to form giving rise to inhomogeneous superconductivity.  

The temperature-dependent Hall resistance exhibits a linear dependence on the magnetic field, B, as shown in Inset Figure \ref{fig:transport1}C. A change in slope and, consequently, the sign of the Hall coefficient occurs at 65 K. This change indicates a shift from electron-dominated conduction at low temperatures to hole-dominated conduction at higher temperatures, suggesting the coexistence of both electron and hole pockets at the Fermi level, in agreement with previously reported theoretical predictions \cite{shao2016spin}. The sheet carrier density and mobility at 10 K are 1.6 $\times$ 10$^{16}$ cm$^{-2}$ and 1.2 cm$^2$/Vs, respectively.
These carrier density values are consistent with previous observations\cite{qiao2024observation}. The carrier concentrations are 3-4 orders of magnitude higher than observed for the two-dimensional electron gas at SrTiO$_3$ \cite{lomker2019redox, kormondy2018large} and KTaO$_3$ \cite{zou2015latio3,zhang2018high, kumar2021observation, liu2021two} interfaces. Furthermore, Figure \ref{fig:transport1}D compares the resistance as a function of the nominal p-BBO thickness. The number of Eu monolayers is tuned to match the p-BBO thicknesses. The transition temperatures are comparable and independent of the nominal p-BBO thicknesses.

\begin{figure*} 
	\centering
	\includegraphics[width=1\textwidth]{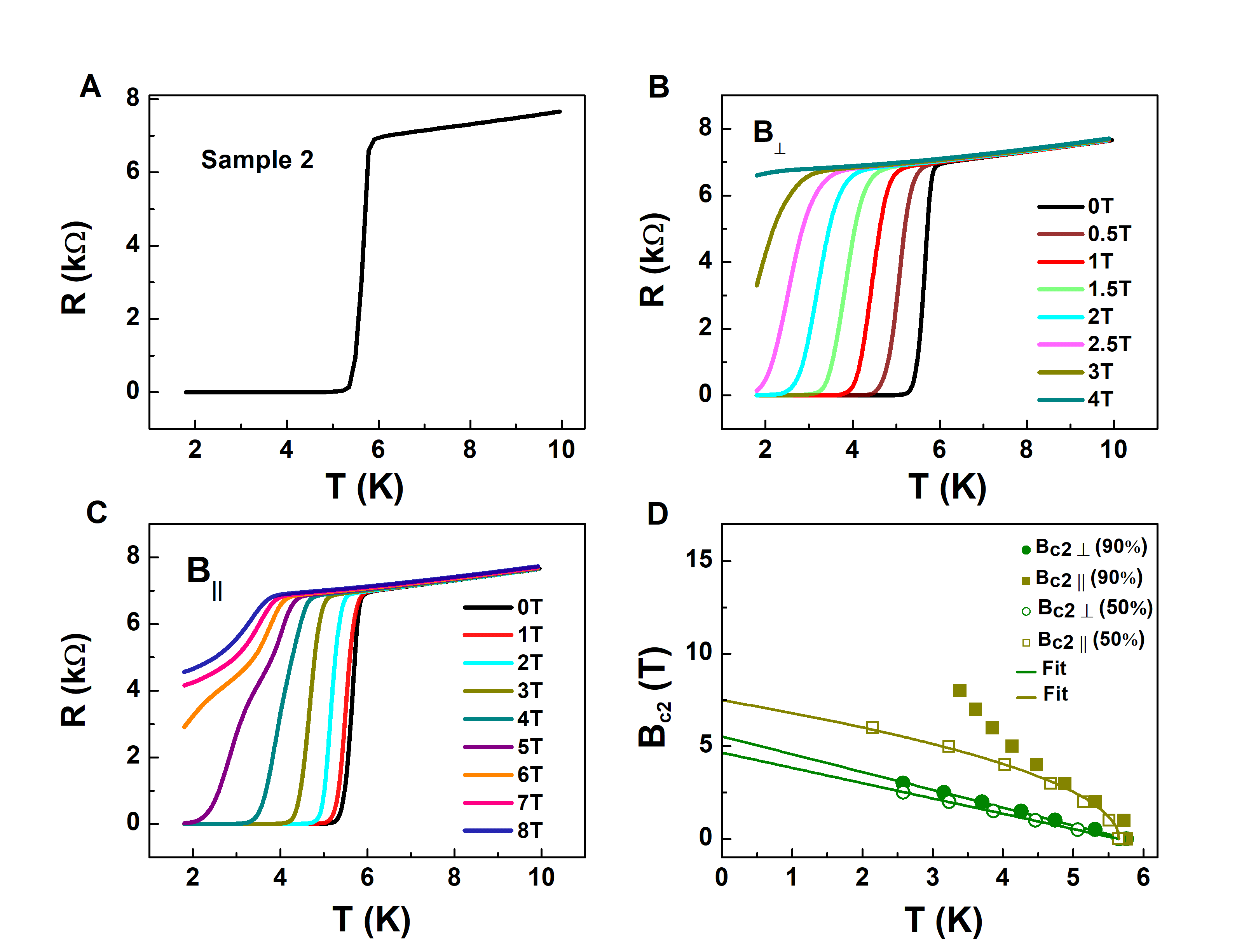} 
	\caption{\textbf{Magnetotransport measurements of Eu/BBOx (Sample 2) on STO (001)} \textbf{(A)} Resistance versus temperature for Sample 2. \textbf{(B)} Magnetoresistance measured as a function of temperature for out-of-plane B configuration. \textbf{(C)} Magnetoresistance measured as a function of temperature for the in-plane B configuration. \textbf{(D)} The upper critical field (B$_{c2}$(T)) as a function of temperature for in-plane and out-of-plane configurations. The upper critical field (B$_{c2}$(T)) were determined using 50 \% and 90 \% of normal-state resistance criterion.  The green and dark yellow solid lines represent the fit using the 2D Ginzburg-Landau (GL) equations for in-plane and out-of-plane configurations: $B_{c2\perp}(T)= B_{c2\perp} (0) \left( 1-\frac{T}{T_c} \right)$ and $B_{c2\parallel}(T)= B_{c2\parallel} (0) \left( 1-\frac{T}{T_c} \right)^{1/2}$.}
	\label{fig:transport2} 
\end{figure*}

To further investigate the dimensionality of the system, magnetotransport measurements were performed on a second Eu/p-BBO  sample (Sample 2). The zero-field temperature-dependent resistance is presented in Figure \ref{fig:transport2}A, with the onset of superconductivity, $T_{c, onset}$, observed at 5.76 K. Temperature-dependent magnetoresistance measurements were carried out with the magnetic field applied in the out-of-plane direction perpendicular to the current direction, $B_\perp$ (Figure  \ref{fig:transport2}B), or in-plane direction parallel to the current direction, $B_{||}$ (Figure  \ref{fig:transport2}C). $B_\perp$ significantly suppresses superconductivity, with the transition nearly vanishing at B = 4 T. In contrast, for $B_{||}$, superconductivity persists at B = 8 T, as shown in Figure \ref{fig:transport2}C. The anisotropic dependence of $T_c$ on the direction of the magnetic field suggests quasi-two-dimensional superconductivity.

The critical fields for both out-of-plane, $B_{c2\perp}$, and in-plane $B_{c2\parallel}$ configurations were determined using the 90\% and 50\% of normal resistance criterion. $B_{c2\perp}$ and $B_{c2\parallel}$ are plotted as a function of temperature in Figure \ref{fig:transport2}D. To determine the upper critical fields at T = 0 K, the data were fitted using the two-dimensional Ginzburg-Landau (GL) model for the out-of-plane configuration and in-plane configuration \cite{tinkham1996introduction}:

\begin{equation}
B_{c2\perp}(T)= B_{c2\perp}(0) \left(1-\frac{T}{T_c}\right)
\end{equation}

and 
\begin{equation}
 B_{c2\parallel}(T)= B_{c2\parallel}(0)\left(1-\frac{T}{T_c} \right)^{\frac{1}{2}} 
\end{equation}

where $B_{c2\perp}(0)$ and $B_{c2\parallel}(0)$ represent the extrapolated upper critical fields at 0 K. The extracted values of $B_{c2\perp}(0)$ using the 90\% and 50\% criteria are 5.54 T and 4.66 T, respectively, both exceeding the reported $B_{c2\perp}(0)$ value of 2.2 T for bulk BaBi$_3$ single crystals \cite{haldolaarachchige2014superconducting, lu2015evidence}. As shown in Figure \ref{fig:transport2}D, the temperature dependence of the in-plane upper critical field $B_{c2\parallel}$ determined using the 90\% criterion deviates from the two-dimensional (2D) Ginzburg-Landau (GL) model, whereas $B_{c2\parallel}$ determined by the 50\% criterion follows the expected 2D-GL behavior. The extrapolated value of $B_{c2\parallel}(0)$(50\%) is 7.49 T. These critical fields satisfy the GL relations $B_{c2\perp}(0) = \phi_{0}/2\pi \xi_{GL}^{2} $ and $B_{c2\parallel}(0) = \sqrt12 \phi_{0}/2\pi \xi_{GL} d_{sc}$, where $\phi_{0}$ is the magnetic-flux quantum, $\xi_{GL}$ is the coherence length, and $d_{sc}$ is the effective superconducting thickness. Using the extracted upper critical-field values (50\% criterion),  we obtain $\xi_{GL} = 8.4$ nm, smaller than $d_{sc} = 18.5 $ nm, consistent with quasi-two-dimensional behavior reported for Eu/BBO heterostructures \cite{qiao2024observation}.
 The difference between the in-plane critical fields derived from the 50\% and 90\% criteria originates from a multistage superconducting transition feature appearing near 5 T (Figure \ref{fig:transport2}C). The anisotropy of in-plane and out-of-plane upper critical fields has been reported in various two-dimensional superconducting systems, including Nd$_{1-x}$Eu$_x$NiO$_2$, infinite-layer lanthanum nickelates, LaMnO$_3$/KTaO$_3$, and NbSe$_2$ \cite{wei2023superconducting,  al2023enhanced,zhang2022tailored, sun2023evidence}. 

 Recent studies have demonstrated that strong spin-orbit coupling (SOC) in two-dimensional superconductors can lead to out-of-plane spin locking and the formation of a uniaxial Ising spin texture\cite{lu2015evidence, saito2016superconductivity}. This spin-momentum locking suppresses Zeeman depairing, a key characteristic of Ising superconductivity, and can result in highly anisotropic superconducting behavior, with in-plane upper critical fields exceeding the Pauli limit. In contrast, Rashba-type SOC leads to in-plane spin locking, which enhances in-plane $B_{c2||}$ to approximately $\sqrt{2}$ B$_P$\cite{Rashba2001}. Furthermore, spin-orbit scattering can randomize electron spins, reducing paramagnetic limiting effects and thereby increasing $B_{c2||}$ \cite{SOS1962}. Notably, the Pauli paramagnetic limit, given by $B_P = 1.84~ T_c (B=0)$, is about 10.6 T for the BaBi$_3$ films \cite{chandrasekhar1962note, clogston1962upper}. While the extracted value of   $B_{c2\parallel}(0)$ (50 \%) = 7.49 T is below this Pauli limit, Figure \ref{fig:transport2}C shows that the superconducting state is not fully suppressed for $B_{\parallel}=$ 8 T and may approach the Pauli limit at 0 K.
 
 \begin{figure*} 
 	\centering
 	\includegraphics[width=1\textwidth]{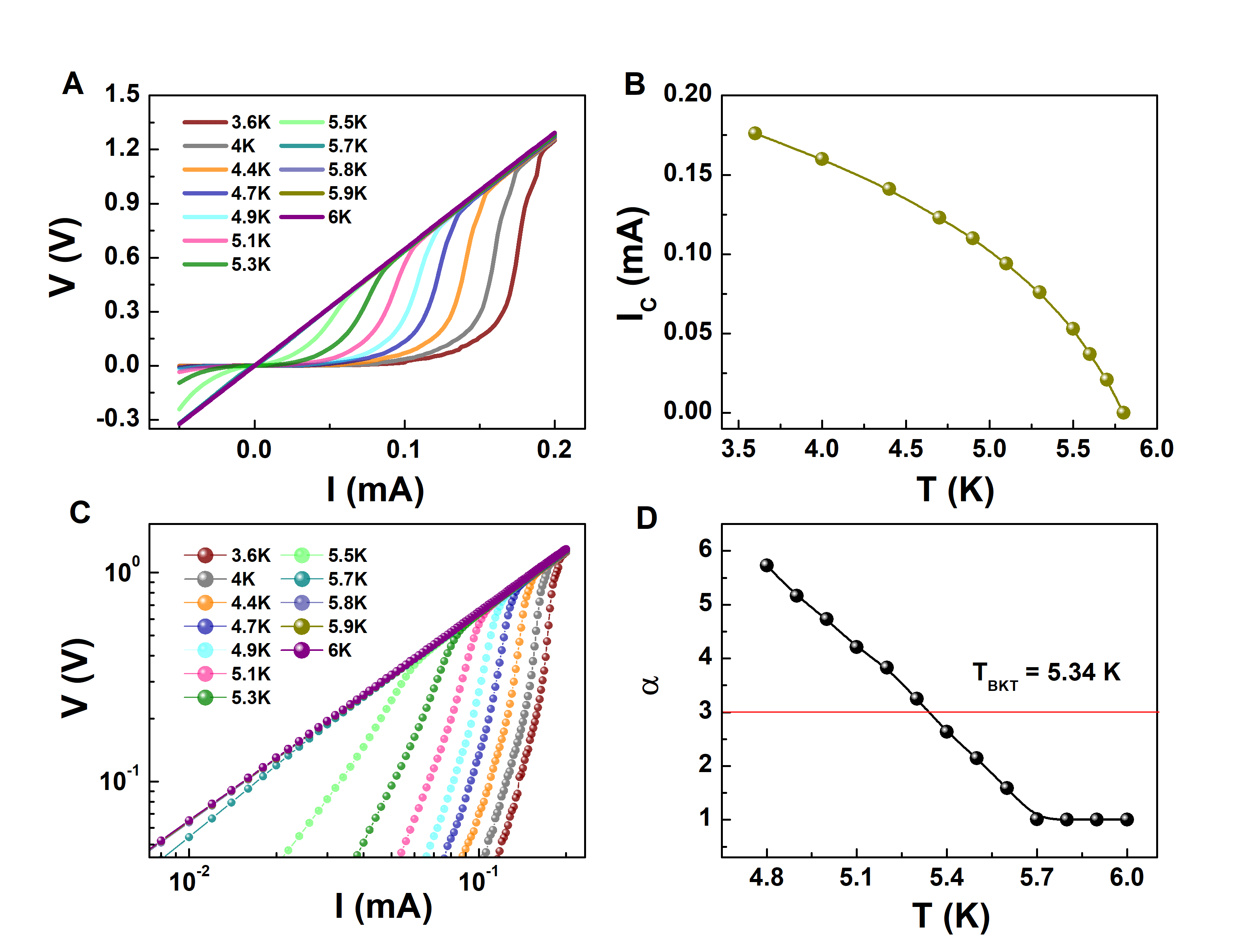} 
 
 	\caption{\textbf{ Current-Voltage (I-V) measurements of EuO/BBOx (Sample 2) on STO (001)} \textbf{(A)} I-V characteristics measured at various temperatures. \textbf{(B)} Critical current (I$_{c}$), determined from the derivative of the I-V characteristics, plotted as a function of temperature, demonstrating superconducting behavior. \textbf{(C)} I-V curves plotted on a logarithmic scale, to highlight non-linear behaviour more distinctly. \textbf{(D)} Exponent $\alpha$ (extracted from (C)) as a function of temperature: The horizontal solid line indicates the Berezinskii-Kosterlitz-Thouless transition temperature $T_{BKT}\approx$ 5.34 K for $\alpha = 3$. The solid black line is a guide to the eye.}
 	\label{fig:transportIV} 
\end{figure*}

Figure \ref{fig:transportIV}A presents the current-voltage (I-V) characteristic curves as a function of temperature for Sample 2. From the I-V measurements, the critical currents ($I{_c}$) are determined and plotted in Figure \ref{fig:transportIV}B. $I_c$ decreases with increasing temperature and eventually reaches zero at the transition temperature, as expected for a superconductor. We observe that the slope of I-V characteristics transitions from a normal ohmic state ($V \propto I$) to a non-linear power law ($V \propto I^\alpha$) at the superconducting transition temperature. In 2D superconductors, this transition can be understood in the framework of the Berezinskii-Kosterlitz-Thouless (BKT) transition, where the emergence of nonlinear I-V characteristics in the superconducting state results from the current-driven unbinding of vortex-antivortex pairs, which are generated by thermal fluctuations at finite temperatures \cite{beasley1979possibility, epstein1981vortex}. The I-V characteristics are also plotted on a logarithmic scale in Figure \ref{fig:transportIV}C, highlighting the non-linearity more distinctly. To further analyze this transition, we extracted the exponent $\alpha$ from the I-V curves and plot $\alpha$ as a function of temperature in Figure \ref{fig:transportIV}D. The BKT transition temperature ($T_{BKT}$) is estimated to be around 5.34 K by interpolating to $\alpha = 3$, which closely matches $T_{c,0}$ obtained from resistance measurements. Moreover, the temperature-dependent resistance near T$_{BKT}$ is expected to follow the Halperin-Nelson (HN) formula: R = R$_{0}$ exp(-bt$^{-1/2}$ ), where R$_{0}$ and b are material-specific parameters and t = T/T$_{BKT}$ - 1  \cite{HNfit1979}. Figure S5 (Supplementary Materials\cite{supplement}) presents the fit based on the HN formula, yielding T$_{BKT}$ = 5.31 K, which agrees well with the value obtained from the I-V analysis.  The BKT transition is also measured in a third sample in Figure S6 in the Supplementary Materials.\cite{supplement} This agreement with BKT theory is consistent with two-dimensional (2D) nature of superconductivity in the heterostructures. 
 
\begin{figure*} 
	\centering
	\includegraphics[width=1\textwidth]{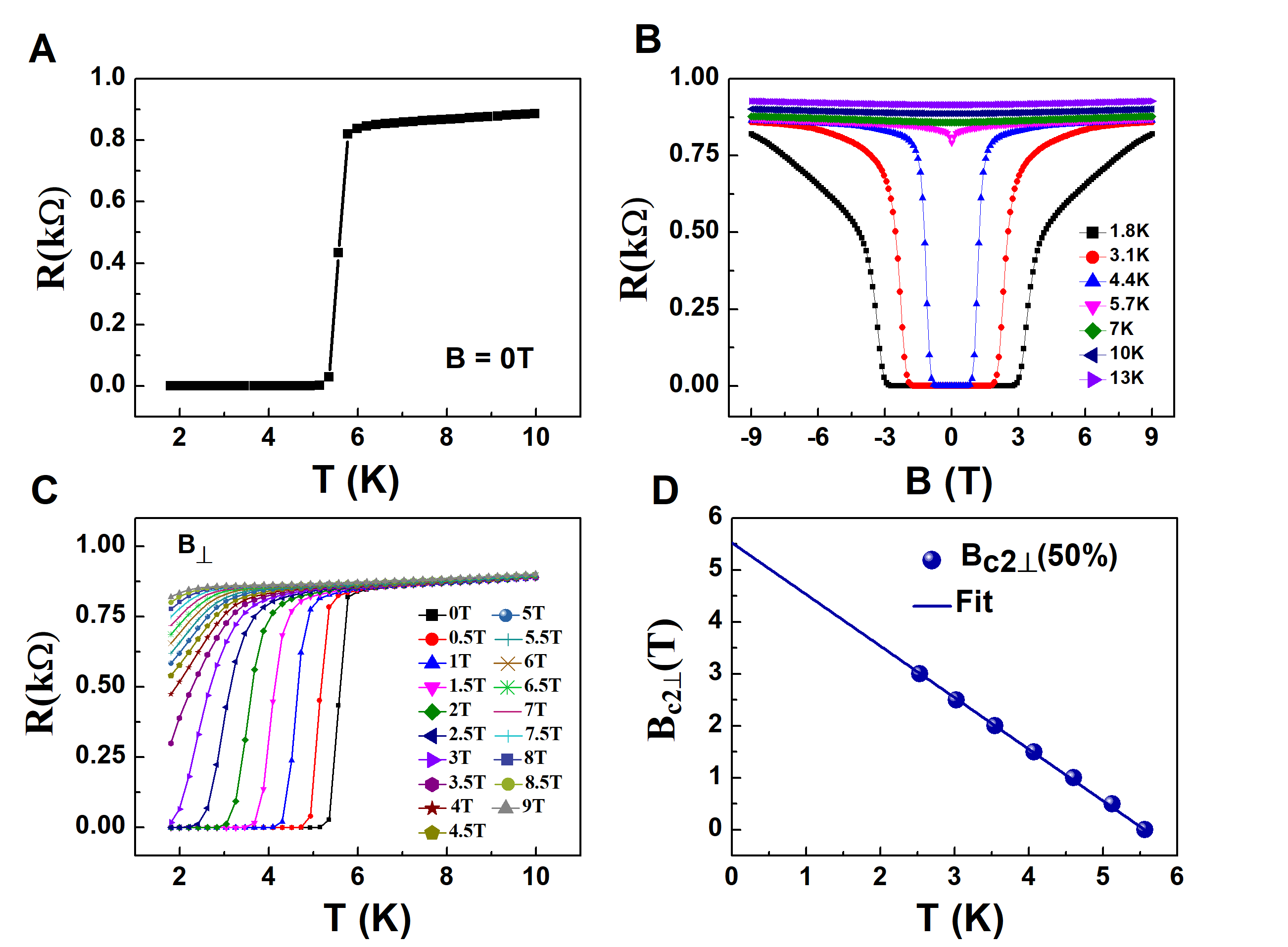} 

	\caption{\textbf{Magnetotransport measurements of nominal Al/p-BBO on STO (001)} \textbf{(A)} Resistance versus temperature.\textbf{(B)} Magnetoresistance measured at various temperatures above and below the superconducting transition. \textbf{(C)} Resistance measured as a function of temperature for out-of-plane B configuration. \textbf{(D)} The temperature dependence of the upper critical field (B$_{c2\perp}$(T)) for out-of-plane configuration. The solid line represents the fit using the two-dimensional Ginzburg-Landau (GL) model, given by $B_{c2\perp}(T)= B_{c2\perp} (0) \left( 1-\frac{T}{T_c} \right)$.} 
	\label{fig:transport3} 
\end{figure*}

To examine the role of the capping layer, we performed magnetotransport measurements on nominal Al/p-BBO heterostructures. Figure \ref{fig:transport3}A presents the temperature-dependent resistance, which shows superconducting behaviour with a transition temperature of $\approx$ 5.75 K, comparable to that observed in Eu/p-BBO. Magnetoresistance curves measured with the out-of-plane perpendicular magnetic field (B$_\perp$) at various temperatures are shown in Figure \ref{fig:transport3}B. Figure \ref{fig:transport3}C shows the resistance as a function of temperature for different values of $B_\perp$ with a decrease in $T_c$ observed with increasing $B_\perp$. The out-of-plane critical field, $B_{c2\perp}$, extracted using the 50\% normal resistance criterion, is plotted as a function of temperature in Figure \ref{fig:transport3}D. The value of $B_{c2\perp}(0)$ extracted by fitting the 2D-GL equation for the out-of-plane configuration is 5.52 T, which is close to the values observed in the Eu/p-BBO samples. These results indicate that the choice of capping layer has little influence on the superconducting properties of BaBi$_3$. 

Based on the structural transitions evidenced by the XRD and STEM measurements, the emergent superconductivity in the nominal Eu/p-BBO and Al/p-BBO heterostructures arises from the formation of BaBi$_3$ via the reduction of p-BBO. The structural results presented here rule out an electronic reconstruction at the interface between EuO and oxygen-deficient perovskite BaBiO$_{3-x}$ and/or electron doping of BaBiO$_{3-x}$ as the origin of superconductivity in nominal Eu/p-BBO heterostructures\cite{qiao2024observation}.
Studies of oxygen-deficient perovskite BaBiO$_{3-x}$ find an enhanced electronic band gap, suggesting that electron doping via oxygen vacancies may not stabilize superconductivity \cite{cao2021realization, lightfoot1991babio2}. We also consider the possibility of the reduction of p-BBO leading to the formation of an infinite-layered bismuthate structure analogous to the superconducting infinite-layer structures, which are also stabilized by the reduction of the perovskite phase\cite{li2019superconductivity, hepting2020electronic, wei2023superconducting, shengwei2022superconductivity, nomura2022superconductivity}. First-principle calculations for an infinite-layer type BaBiO$_2$ have recently been reported with a predicted T$_c$ of 3.9 - 7.7 K under pressure and with spin-orbit coupling\cite{cui2023spin}. Here, superconductivity emerges on suppression of rotation of the oxygen sublattice at 10 GPa. However, the lattice parameters of the proposed structure are incompatible with our experimental results. It is however, plausible that the layered infinite-layer phase forms as an intermediate phase and decomposes into BaO and BaBi$_3$. For example, bulk stoichiometric NdNiO$_2$ is known to be thermodynamically unstable and decomposes into Nd$_2$O$_3$, NiO, and Ni\cite{malyi2022bulk}.

The anisotropic critical magnetic fields and the presence of a BKT transition are consistent with quasi-two-dimensional superconductivity.\cite{qiao2024observation} However, the crystal structure of BaBi$_3$ is inherently three-dimensional, and the nominal thicknesses of the p-BBO films (13-36 nm) exceed the superconducting coherence length, $\xi_{GL} = 8.4$ nm. In addition, structural and chemical disorder arising from phase separation between BaO and BaBi$_3$ may lead to spatially inhomogeneous superconductivity. Another relevant length scale for assessing dimensionality is the superconducting London penetration depth, $\lambda_L(0)$. For bulk BaBi$_3$, $\lambda_L(0) \approx 130$ nm, which is an order of magnitude larger than the film thickness $d$.\cite{haberkorn2023superconducting, haldolaarachchige2014superconducting} Therefore, in the limit $d \ll \lambda_L(0)$, the superconductivity can be effectively regarded as two-dimensional. Future studies should aim to unravel the microscopic mechanisms governing the quasi-two-dimensional superconducting state and quantify the extent to which disorder modulates these phenomena.

\subsection*{Conclusion}
In conclusion, we observe superconductivity at 6 K in \textit{in situ} reduced perovskite BaBiO$_3$ films. The structural characterization of the superconducting films points to the decomposition of the reduced perovskite phase to form the superconducting BaBi$_3$ phase. Although BaBi$_3$ has been studied in bulk crystals due to the role of spin-orbit coupling in enhancing superconductivity, realizing superconductivity in films with reduced dimensionality allows investigating the effects of confinement and proximity effects with other bismuth-based layered topological quantum materials\cite{xia2009observation, murakami2006quantum}.


\section*{Acknowledgments}
D.P.K. and S. acknowledge support from the NSF DMR 2324174 and the Duke Endowment Duke Science and Technology Fund. Transport measurements by J.T.M. and G.F. were supported by the National Science Foundation grant DMR-2327535. STEM imaging was performed at the Analytical Instrumentation Facility (AIF) at North Carolina State University, which is supported by the State of North Carolina and the National Science Foundation (award number ECCS-2025064). The AIF is a member of the North Carolina Research Triangle Nanotechnology Network (RTNN), a site in the National Nanotechnology Coordinated Infrastructure (NNCI).

\paragraph*{Competing interests:}
There are no competing interests to declare.

\paragraph*{Data and materials availability:}
All data are available in the main text or the supplementary materials.

%


\begin{thebibliography}{10}
\providecommand{\url}[1]{\texttt{#1}}
\expandafter\ifx\csname urlstyle\endcsname\relax
  \providecommand{\doi}[1]{doi:\discretionary{}{}{}#1}\else
  \providecommand{\doi}{doi:\discretionary{}{}{}\begingroup \urlstyle{rm}\Url}\fi

\bibitem{sleight1975high}
A.~W. Sleight, J.~Gillson, P.~Bierstedt, High-temperature superconductivity in the {BaPb$_{1-x}$Bi$_x$O$_3$} systems. \emph{Solid State Communications} \textbf{17}~(1), 27--28 (1975).

\bibitem{mattheiss1988superconductivity}
L.~Mattheiss, E.~Gyorgy, D.~Johnson~Jr, {Superconductivity above 20 K in the Ba-K-Bi-O system}. \emph{Physical Review B} \textbf{37}~(7), 3745 (1988).

\bibitem{cava1988superconductivity}
R.~Cava, B.~Batlogg, J.~Krajewski, R.~Farrow, L.~Rupp~Jr, A.~White, K.~Short, W.~Peck, T.~Kometani, Superconductivity near 30 K without copper: the {Ba$_{0.6}$K$_{0.4}$BiO$_3$} perovskite. \emph{Nature} \textbf{332}~(6167), 814--816 (1988).

\bibitem{hinks1988synthesis}
D.~Hinks, B.~Dabrowski, J.~Jorgensen, A.~Mitchell, D.~Richards, S.~Pei, D.~Shi, Synthesis, structure and superconductivity in the {Ba$_{1- x}$ K$_x$BiO$_{3-y}$} system. \emph{Nature} \textbf{333}~(6176), 836--838 (1988).

\bibitem{pei1990structural}
S.~Pei, J.~Jorgensen, B.~Dabrowski, D.~Hinks, D.~Richards, A.~Mitchell, J.~Newsam, S.~Sinha, D.~Vaknin, A.~Jacobson, Structural phase diagram of the {Ba$_{1-x}$K$_x$BiO$_3$} system. \emph{Physical Review B} \textbf{41}~(7), 4126 (1990).

\bibitem{cox1976crystal}
D.~Cox, A.~Sleight, Crystal structure of {Ba$_2$Bi$^{3+}$Bi$^{5+}$O$^6$}. \emph{Solid State Communications} \textbf{19}~(10), 969--973 (1976).

\bibitem{sleight2015bismuthates}
A.~W. Sleight, Bismuthates: {BaBiO$_{3}$} and related superconducting phases. \emph{Physica C: Superconductivity and its Applications} \textbf{514}, 152--165 (2015).

\bibitem{foyevtsova2015hybridization}
K.~Foyevtsova, A.~Khazraie, I.~Elfimov, G.~A. Sawatzky, Hybridization effects and bond disproportionation in the bismuth perovskites. \emph{Physical Review B} \textbf{91}~(12), 121114 (2015).

\bibitem{shao2016spin}
D.~Shao, X.~Luo, W.~Lu, L.~Hu, X.~Zhu, W.~Song, X.~Zhu, Y.~Sun, Spin-orbit coupling enhanced superconductivity in Bi-rich compounds {$ABi_3$ (A= Sr and Ba)}. \emph{Scientific Reports} \textbf{6}~(1), 21484 (2016).

\bibitem{powell2025multiphase}
L.~Powell, W.~Kuang, G.~Hawkins-Pottier, R.~Jalil, J.~Birkbeck, Z.~Jiang, M.~Kim, Y.~Zou, S.~Komrakova, S.~Haigh, \emph{et~al.}, Multiphase superconductivity in {PdBi$_2$}. \emph{Nature Communications} \textbf{16}~(1), 291 (2025).

\bibitem{matthias1952search}
B.~Matthias, J.~Hulm, A search for new superconducting compounds. \emph{Physical Review} \textbf{87}~(5), 799 (1952).

\bibitem{haldolaarachchige2014superconducting}
N.~Haldolaarachchige, S.~Kushwaha, Q.~Gibson, R.~Cava, Superconducting Properties of {BaBi$_3$}. \emph{Superconductor Science and Technology} \textbf{27}~(10), 105001 (2014).

\bibitem{jha2016hydrostatic}
R.~Jha, M.~A. Avila, R.~A. Ribeiro, Hydrostatic pressure effect on the superconducting properties of {BaBi$_3$} and {SrBi$_3$} single crystals. \emph{Superconductor Science and Technology} \textbf{30}~(2), 025015 (2016).

\bibitem{wang2021superconducting}
Y.~Wang, T.~Taguchi, H.~Li, A.~Suzuki, Y.~Zhang, A.~Miura, M.~Ikeda, H.~Goto, R.~Eguchi, T.~Miyazaki, \emph{et~al.}, Superconducting properties of {BaBi$_3$} at ambient and high pressures. \emph{Physical Chemistry Chemical Physics} \textbf{23}~(40), 23014--23023 (2021).

\bibitem{haberkorn2023superconducting}
N.~Haberkorn, R.~A. Ribeiro, L.~Xiang, S.~Bud'ko, P.~C. Canfield, Superconducting properties and vortex pinning in intermetallic {BaBi$_3$} single crystals: a magnetization study. \emph{Physica C: Superconductivity and its Applications} \textbf{615}, 1354387 (2023).

\bibitem{wang2018surface}
W.-L. Wang, Y.-M. Zhang, N.-N. Luo, J.-Q. Fan, C.~Liu, Z.-Y. Dou, L.~Wang, W.~Li, K.~He, C.-L. Song, \emph{et~al.}, Surface symmetry breaking and disorder effects on superconductivity in perovskite {BaBi$_3$} epitaxial films. \emph{Physical Review B} \textbf{98}~(6), 064511 (2018).

\bibitem{zhang2022tailored}
H.~Zhang, A.~Rousuli, K.~Zhang, L.~Luo, C.~Guo, X.~Cong, Z.~Lin, C.~Bao, H.~Zhang, S.~Xu, \emph{et~al.}, Tailored Ising superconductivity in intercalated bulk {NbSe$_2$}. \emph{Nature Physics} \textbf{18}~(12), 1425--1430 (2022).

\bibitem{li2015topological}
G.~Li, B.~Yan, R.~Thomale, W.~Hanke, Topological nature and the multiple Dirac cones hidden in Bismuth {high-T$_c$} superconductors. \emph{Scientific Reports} \textbf{5}~(1), 10435 (2015).

\bibitem{yan2013large}
B.~Yan, M.~Jansen, C.~Felser, A large-energy-gap oxide topological insulator based on the superconductor {BaBiO$_{3}$}. \emph{Nature Physics} \textbf{9}~(11), 709--711 (2013).

\bibitem{zapf2018domain}
M.~Zapf, M.~St{\"u}binger, L.~Jin, M.~Kamp, F.~Pfaff, A.~Lubk, B.~B{\"u}chner, M.~Sing, R.~Claessen, Domain matching epitaxy of {BaBiO$_{3}$} on {SrTiO$_3$} with structurally modified interface. \emph{Applied Physics Letters} \textbf{112}~(14) (2018).

\bibitem{qiao2024observation}
W.~Qiao, J.~Zhao, Y.~Chen, S.~Cao, W.~Xing, R.~Cai, L.~Guo, T.~Qian, X.~Xie, W.~Han, Observation of quasi-two-dimensional superconductivity at the {EuO}-{BaBiO$_{3}$} interface. \emph{Physical Review B} \textbf{109}~(5), 054509 (2024).

\bibitem{guo2018euo}
W.~Guo, A.~B. Posadas, S.~Lu, D.~J. Smith, A.~A. Demkov, EuO epitaxy by oxygen scavenging on {SrTiO$_3$} (001): Effect of {SrTiO$_3$} thickness and temperature. \emph{Journal of Applied Physics} \textbf{124}~(23), 235301 (2018).

\bibitem{mairoser2015high}
T.~Mairoser, J.~A. Mundy, A.~Melville, D.~Hodash, P.~Cueva, R.~Held, A.~Glavic, J.~Schubert, D.~A. Muller, D.~G. Schlom, \emph{et~al.}, High-quality {EuO} thin films the easy way via topotactic transformation. \emph{Nature Communications} \textbf{6}~(1), 7716 (2015).

\bibitem{averyanov2015direct}
D.~V. Averyanov, Y.~G. Sadofyev, A.~M. Tokmachev, A.~E. Primenko, I.~A. Likhachev, V.~G. Storchak, Direct epitaxial integration of the ferromagnetic semiconductor EuO with silicon for spintronic applications. \emph{{ACS Appl. Mat. \& Interfaces}} \textbf{7}~(11), 6146--6152 (2015).

\bibitem{zollweg1955x}
R.~J. Zollweg, X-ray lattice constant of barium oxide. \emph{Physical Review} \textbf{100}~(2), 671 (1955).

\bibitem{matthias1961ferromagnetic}
B.~Matthias, R.~Bozorth, J.~Van~Vleck, Ferromagnetic interaction in {EuO}. \emph{Physical Review Letters} \textbf{7}~(5), 160 (1961).

\bibitem{supplement}
See Supplemental Material at [url] for additional experimental results.

\bibitem{cao2021realization}
H.~Cao, H.~Guo, Y.-C. Shao, Q.~Liu, X.~Feng, Q.~Lu, Z.~Wang, A.~Zhao, A.~Fujimori, Y.-D. Chuang, \emph{et~al.}, Realization of electron antidoping by modulating the breathing distortion in {BaBiO$_{3}$}. \emph{Nano Letters} \textbf{21}~(9), 3981--3988 (2021).

\bibitem{wang2018two}
B.~Wang, X.~Luo, K.~Ishigaki, K.~Matsubayashi, J.~Cheng, Y.~Sun, Y.~Uwatoko, Two distinct superconducting phases and pressure-induced crossover from type-II to type-I superconductivity in the spin-orbit-coupled superconductors BaBi$_3$ and SrBi$_3$. \emph{Phys. Rev. B} \textbf{98}~(22), 220506 (2018).

\bibitem{zverev2009transport}
V.~Zverev, A.~Korobenko, G.~Sun, D.~Sun, C.~Lin, A.~Boris, Transport properties and the anisotropy of Ba1- x K x Fe2As2 single crystals in normal and superconducting states. \emph{JETP letters} \textbf{90}~(2), 130--133 (2009).

\bibitem{Yuan2022strange}
J.~Yuan, Q.~Chen, K.~Jiang, Z.~Feng, Z.~Lin, H.~Yu, G.~He, J.~Zhang, X.~Jiang, X.~Zhang, Y.~Shi, Y.~Zhang, M.~Qin, Z.~G. Cheng, N.~Tamura, Y.-f. Yang, T.~Xiang, J.~Hu, I.~Takeuchi, K.~Jin, Z.~Zhao, {Scaling of the strange-metal scattering in unconventional superconductors}. \emph{Nature} \textbf{602}, 431–436 (2022).

\bibitem{saito2016superconductivity}
Y.~Saito, Y.~Nakamura, M.~S. Bahramy, Y.~Kohama, J.~Ye, Y.~Kasahara, Y.~Nakagawa, M.~Onga, M.~Tokunaga, T.~Nojima, \emph{et~al.}, Superconductivity protected by spin--valley locking in ion-gated {MoS$_2$}. \emph{Nature Physics} \textbf{12}~(2), 144--149 (2016).

\bibitem{chen2021superconductivity}
Z.~Chen, Z.~Liu, Y.~Sun, X.~Chen, Y.~Liu, H.~Zhang, H.~Li, M.~Zhang, S.~Hong, T.~Ren, C.~Zhang, H.~Tian, Y.~Zhou, J.~Sun, Y.~Xie, Two-Dimensional Superconductivity at the ${\mathrm{LaAlO}}_{3}/{\mathrm{KTaO}}_{3}(110)$ Heterointerface. \emph{Phys. Rev. Lett.} \textbf{126}, 026802 (2021).

\bibitem{abrikosov2017fundamentals}
A.~A. Abrikosov, \emph{Fundamentals of the Theory of Metals} (NorthHolland, Amsterdam) (1988).

\bibitem{jaeger1989onset}
H.~Jaeger, D.~Haviland, B.~Orr, A.~Goldman, Onset of superconductivity in ultrathin granular metal films. \emph{Physical Review B} \textbf{40}~(1), 182 (1989).

\bibitem{lomker2019redox}
P.~L{\"o}mker, M.~M{\"u}ller, Redox-controlled epitaxy and magnetism of oxide heterointerfaces: {EuO/SrTiO$_3$}. \emph{Physical Review Materials} \textbf{3}~(6), 061401 (2019).

\bibitem{kormondy2018large}
K.~J. Kormondy, L.~Gao, X.~Li, S.~Lu, A.~B. Posadas, S.~Shen, M.~Tsoi, M.~R. McCartney, D.~J. Smith, J.~Zhou, \emph{et~al.}, Large positive linear magnetoresistance in the two-dimensional t2g electron gas at the {EuO/SrTiO$_3$} interface. \emph{Scientific Reports} \textbf{8}~(1), 7721 (2018).

\bibitem{zou2015latio3}
K.~Zou, S.~Ismail-Beigi, K.~Kisslinger, X.~Shen, D.~Su, F.~Walker, C.~Ahn, {LaTiO$_3$/KTaO$_3$} interfaces: A new two-dimensional electron gas system. \emph{APL Materials} \textbf{3}~(3) (2015).

\bibitem{zhang2018high}
H.~Zhang, Y.~Yun, X.~Zhang, H.~Zhang, Y.~Ma, X.~Yan, F.~Wang, G.~Li, R.~Li, T.~Khan, \emph{et~al.}, High-mobility spin-polarized two-dimensional electron gases at{ {EuO/KTaO$_3$}} interfaces. \emph{Physical Review Letters} \textbf{121}~(11), 116803 (2018).

\bibitem{kumar2021observation}
N.~Kumar, N.~Wadehra, R.~Tomar, Shama, S.~Kumar, Y.~Singh, S.~Dattagupta, S.~Chakraverty, Observation of shubnikov-de haas oscillations, planar hall effect, and anisotropic magnetoresistance at the conducting interface of {EuO--KTaO$_3$}. \emph{Advanced Quantum Technologies} \textbf{4}~(1), 2000081 (2021).

\bibitem{liu2021two}
C.~Liu, X.~Yan, D.~Jin, Y.~Ma, H.-W. Hsiao, Y.~Lin, T.~M. Bretz-Sullivan, X.~Zhou, J.~Pearson, B.~Fisher, \emph{et~al.}, Two-dimensional superconductivity and anisotropic transport at {KTaO$_3$} (111) interfaces. \emph{Science} \textbf{371}~(6530), 716--721 (2021).

\bibitem{tinkham1996introduction}
M.~Tinkham, {Introduction to superconductivity, Mac-Graw-Hill Inc}. \emph{New York, London, Tokyo}  (1996).

\bibitem{lu2015evidence}
J.~Lu, O.~Zheliuk, I.~Leermakers, N.~F. Yuan, U.~Zeitler, K.~T. Law, J.~Ye, Evidence for two-dimensional Ising superconductivity in gated {MoS$_2$}. \emph{Science} \textbf{350}~(6266), 1353--1357 (2015).

\bibitem{wei2023superconducting}
W.~Wei, D.~Vu, Z.~Zhang, F.~J. Walker, C.~H. Ahn, Superconducting {Nd$_{1-x}$Eu$_x$NiO$_2$} thin films using in situ synthesis. \emph{Science Advances} \textbf{9}~(27), eadh3327 (2023).

\bibitem{al2023enhanced}
A.~H. Al-Tawhid, S.~J. Poage, S.~Salmani-Rezaie, A.~Gonzalez, S.~Chikara, D.~A. Muller, D.~P. Kumah, M.~N. Gastiasoro, J.~Lorenzana, K.~Ahadi, Enhanced critical field of superconductivity at an oxide interface. \emph{Nano Letters} \textbf{23}~(15), 6944--6950 (2023).

\bibitem{sun2023evidence}
W.~Sun, Y.~Li, R.~Liu, J.~Yang, J.~Li, W.~Wei, G.~Jin, S.~Yan, H.~Sun, W.~Guo, \emph{et~al.}, Evidence for Anisotropic Superconductivity Beyond Pauli Limit in Infinite-Layer Lanthanum Nickelates. \emph{Advanced Materials} \textbf{35}~(32), 2303400 (2023).

\bibitem{Rashba2001}
L.~P. Gor'kov, E.~I. Rashba, Superconducting 2D System with Lifted Spin Degeneracy: Mixed Singlet-Triplet State. \emph{Phys. Rev. Lett.} \textbf{87}, 037004 (2001), \doi{10.1103/PhysRevLett.87.037004}, \url{https://link.aps.org/doi/10.1103/PhysRevLett.87.037004}.

\bibitem{SOS1962}
A.~A. ABRIKOSOV, L.~P. GOR'KOV, SPIN-ORBIT INTERACTION AND THE KNIGHT SHIFT IN SUPERCONDUCTORS. \emph{Sov. Phys. JETP} \textbf{15}, 752 (1962).

\bibitem{chandrasekhar1962note}
B.~Chandrasekhar, A note on the maximum critical field of high-field superconductors. \emph{Appl. Phys. Letters} \textbf{1}, 7--8 (1962).

\bibitem{clogston1962upper}
A.~M. Clogston, Upper limit for the critical field in hard superconductors. \emph{Physical Review Letters} \textbf{9}~(6), 266 (1962).

\bibitem{beasley1979possibility}
M.~Beasley, J.~Mooij, T.~Orlando, Possibility of vortex-antivortex pair dissociation in two-dimensional superconductors. \emph{Physical Review Letters} \textbf{42}~(17), 1165 (1979).

\bibitem{epstein1981vortex}
K.~Epstein, A.~Goldman, A.~Kadin, Vortex-antivortex pair dissociation in two-dimensional superconductors. \emph{Physical Review Letters} \textbf{47}~(7), 534 (1981).

\bibitem{HNfit1979}
B.~I. Halperin, D.~R. Nelson, Resistive transition in superconducting films. \emph{J Low Temp Phys} \textbf{36}, 599–616 (1979).

\bibitem{lightfoot1991babio2}
P.~Lightfoot, J.~A. Hriljac, S.~Pei, Y.~Zheng, A.~Mitchell, D.~Richards, B.~Dabrowski, J.~Jorgensen, D.~Hinks, {BaBiO$_{2.5}$}, a new bismuth oxide with a layered structure. \emph{Journal of Solid State Chemistry} \textbf{92}~(2), 473--479 (1991).

\bibitem{li2019superconductivity}
D.~Li, K.~Lee, B.~Y. Wang, M.~Osada, S.~Crossley, H.~R. Lee, Y.~Cui, Y.~Hikita, H.~Y. Hwang, Superconductivity in an infinite-layer nickelate. \emph{Nature} \textbf{572}~(7771), 624--627 (2019).

\bibitem{hepting2020electronic}
M.~Hepting, D.~Li, C.~Jia, H.~Lu, E.~Paris, Y.~Tseng, X.~Feng, M.~Osada, E.~Been, Y.~Hikita, \emph{et~al.}, Electronic structure of the parent compound of superconducting infinite-layer nickelates. \emph{Nature Materials} \textbf{19}~(4), 381--385 (2020).

\bibitem{shengwei2022superconductivity}
S.~Zeng, C.~Li, L.~E. Chow, Y.~Cao, Z.~Zhang, C.~S. Tang, X.~Yin, Z.~S. Lim, J.~Hu, P.~Yang, \emph{et~al.}, Superconductivity in infinite-layer nickelate {La$_{1- x}$Ca$_x$NiO$_2$} thin films. \emph{Science Advances} \textbf{8}~(7), eabl9927 (2022).

\bibitem{nomura2022superconductivity}
Y.~Nomura, R.~Arita, Superconductivity in infinite-layer nickelates. \emph{Reports on Progress in Physics} \textbf{85}~(5), 052501 (2022).

\bibitem{cui2023spin}
Y.~Cui, H.~Gao, Y.~Li, S.~Xu, H.~Wang, W.~Ren, Spin-orbit coupling enhanced electron-phonon superconductivity in infinite-layer {BaBiO$_{2}$}. \emph{AIP Advances} \textbf{13}~(12) (2023).

\bibitem{malyi2022bulk}
O.~I. Malyi, J.~Varignon, A.~Zunger, Bulk NdNiO2 is thermodynamically unstable with respect to decomposition while hydrogenation reduces the instability and transforms it from metal to insulator. \emph{Physical Review B} \textbf{105}~(1), 014106 (2022).

\bibitem{xia2009observation}
Y.~Xia, D.~Qian, D.~Hsieh, L.~Wray, A.~Pal, H.~Lin, A.~Bansil, D.~Grauer, Y.~S. Hor, R.~J. Cava, \emph{et~al.}, Observation of a large-gap topological-insulator class with a single Dirac cone on the surface. \emph{Nature Physics} \textbf{5}~(6), 398--402 (2009).

\bibitem{murakami2006quantum}
S.~Murakami, Quantum spin Hall effect and enhanced magnetic response by spin-orbit coupling. \emph{Physical Review Letters} \textbf{97}~(23), 236805 (2006).

\end{thebibliography}

%
%
%
%
%
%


\section*{Acknowledgments}
D.P.K. and S. acknowledge support from the NSF (Grant No. DMR 2324174) and the Duke Endowment Duke Science and Technology Fund. Transport measurements by J.T.M. and G.F. were supported by the National Science Foundation grant DMR-2327535. STEM imaging was performed at the Analytical Instrumentation Facility (AIF) at North Carolina State University, which is supported by the State of North Carolina and the National Science Foundation (award number ECCS-2025064). The AIF is a member of the North Carolina Research Triangle Nanotechnology Network (RTNN), a site in the National Nanotechnology Coordinated Infrastructure (NNCI).

\paragraph*{Author contributions:}
Conceptualization: DPK
	Methodology: DPK, GF
	Investigation: S, DPK, GJ,  JTM, MB
	Visualization: S, DPK,  JTM
	Supervision: DPK, GF
	Writing—original draft: DPK
	Writing—review and editing: S, DPK, GF, JTM

\paragraph*{Competing interests:}
There are no competing interests to declare.

\paragraph*{Data and materials availability:}
All data are available in the main text or the supplementary materials.


\subsection*{Supplementary materials}
Materials and Methods\\
Supplementary text\\
Figs. S1 to S6 \\


\newpage


\renewcommand{\thefigure}{S\arabic{figure}}
\renewcommand{\thetable}{S\arabic{table}}
\renewcommand{\theequation}{S\arabic{equation}}
\renewcommand{\thepage}{S\arabic{page}}
\setcounter{figure}{0}
\setcounter{table}{0}
\setcounter{equation}{0}
\setcounter{page}{1} 


\begin{center}
\section*{Supplementary Materials }
\author{
	Shama$^{1}$,
	Jordan T. McCourt$^{1}$,
	Merve Baksi$^{1}$,\and
        Gleb Finkelstein$^{1}$, \&
	Divine Kumah$^{1\ast}$\and
    \\
	\small$^{1}$Department of Physics, Duke University, Durham, NC,  27708, U.S.A.\and
    \\
	\small$^\ast$Corresponding author. Email: divine.kumah@duke.edu\and
}
\end{center}

\subsubsection*{This PDF file includes:}
Materials and Methods\\
Figure S1\\
Figure S2\\
Figure S3\\
Figure S4\\
Figure S5\\
Figure S6\\


\newpage


\subsection*{Materials and Methods}

\subsubsection*{Sample growth} 

Thin films of perovskite-phase BaBiO$_3$ were grown to thickness of 60-90 unit cells (uc) on 5 mm × 5 mm SrTiO$_3$ (001) substrates (commercially available) using molecular beam epitaxy (MBE). Before thin film synthesis, the substrates were cleaned at 575$^o$C in an activated radio frequency oxygen plasma at a chamber pressure of 6x10$^{-6}$ Torr for 15 minutes. Bi, Ba, Eu and Al were evaporated from effusion cells, with their growth rates calibrated using a quartz crystal monitor. The p-BBO thin films were grown under the same conditions as the plasma cleaning process. After growth, the films were cooled in the oxygen plasma to 500$^\circ$C. Subsequently, the oxygen source was turned off, and  20-80 ML Eu metal were deposited in vacuum, followed by an additional 5-10 ML of Eu in an oxygen partial pressure of 2x10$^{-9}$ Torr, as measured by a residual gas analyzer. Next, 10 ML of EuO was deposited in an oxygen partial pressure of 1x10$^{-9}$ Torr. The films were then cooled to room temperature for in situ deposition of Al, forming an AlOx capping layer. Finally, the samples were transferred for ex-situ X-ray diffraction and transport measurements.\textbf{ A second set of samples were fabricated without Eu. Al was deposited at 500 $^o$C in vacuum and the samples were subsequently cooled to room temperature to form nominal Al/BaBiO$_3$ bilayers.}

\subsubsection*{Sample characterization}
X-ray diffraction (XRD) and X-ray reflectivity (XRR) measurements were performed using a high-resolution X-ray diffractometer (Panalytical X'Pert Pro) equipped with a Ge (220) monochromator. The incident X-ray energy was fixed at the Cu-K$\alpha$ wavelength of 8.04 keV. Temperature-dependent resistance and magnetotransport measurements were conducted using a Quantum Design Dynacool Physical Property Measurement System (PPMS) in the van der Pauw configuration. Al wires were ultrasonically bonded to the corners of the 5 mm × 5 mm samples to ensure reliable electrical contacts. The superconducting onset temperature ($T_{c,onset}$) is defined as the temperature at which the resistance drops to 90\% of the normal state resistance, while the superconducting zero temperature (T$_{c,0}$) is defined as the temperature at which the resistance reaches zero. 

Cross-sectional STEM samples of the films were prepared using a Ga+ ion beam in a Thermo Scientific Scios 2 dual beam system. A probe aberration-corrected Thermo Fisher Titan 80-300 kV microscope operating at 200 kV was used for STEM HAADF imaging and EDX measurements. 

Transport measurements were performed by conventional four-probe techniques using Al wires ultrasonically bonded to the samples. The measurements were performed in a Quantum Design Dynacool PPMS system.






\begin{figure*} 
	\centering
	\includegraphics[width=0.5\textwidth]{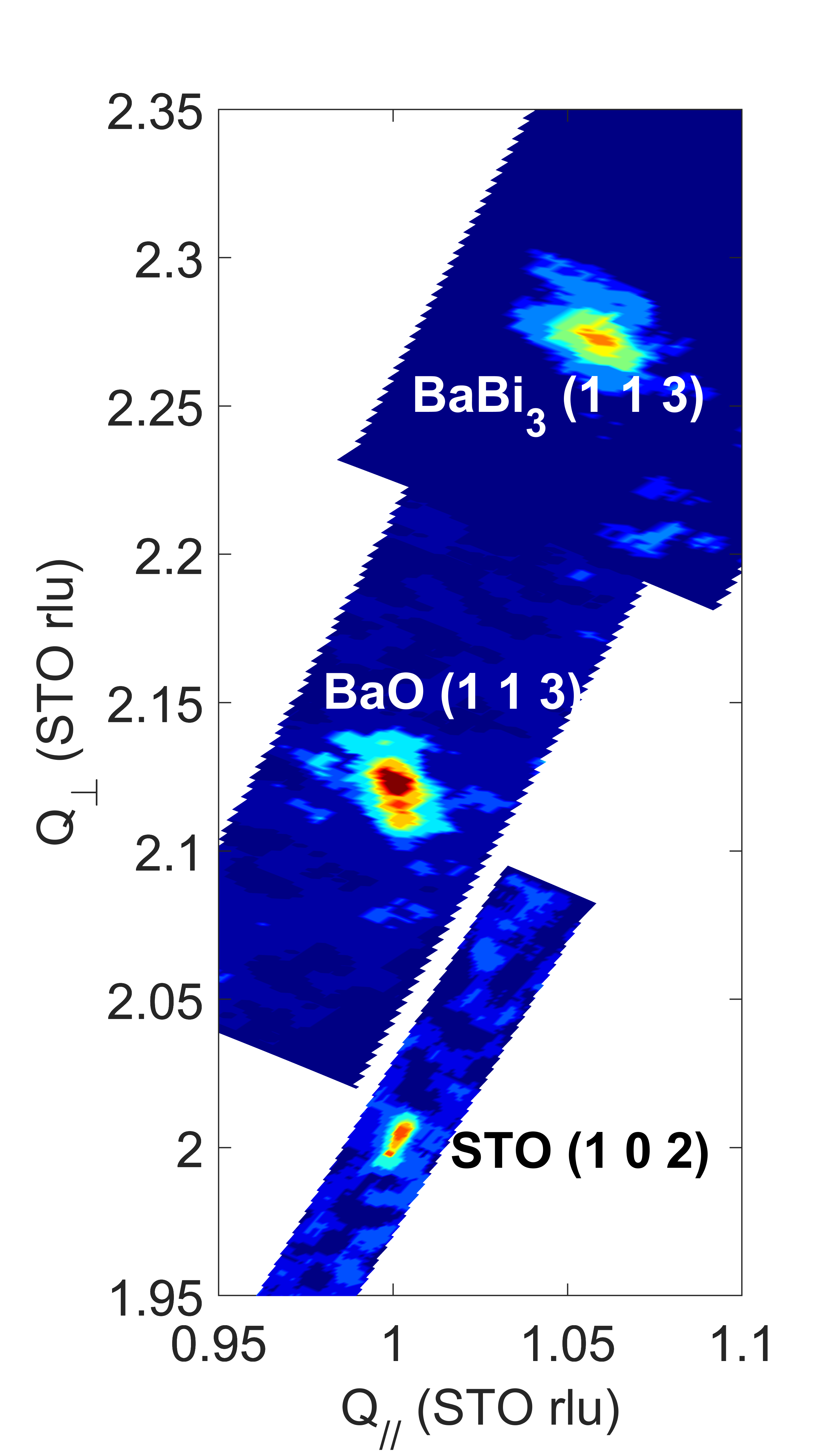} 
	\caption{\textbf{Reciprocal space map of nominal Al/BaBiO$_3$ heterostructure grown on (001)-oriented SrTiO$_3$, showing reflections around the SrTiO$_3$ (001), BaO (1 1 3) and the BaBi$_3$ (1 1 3) Bragg peaks.} }
	\label{fig:RSM} 
\end{figure*}

\begin{figure*} 
	\centering
	\includegraphics[width=1.0\textwidth]{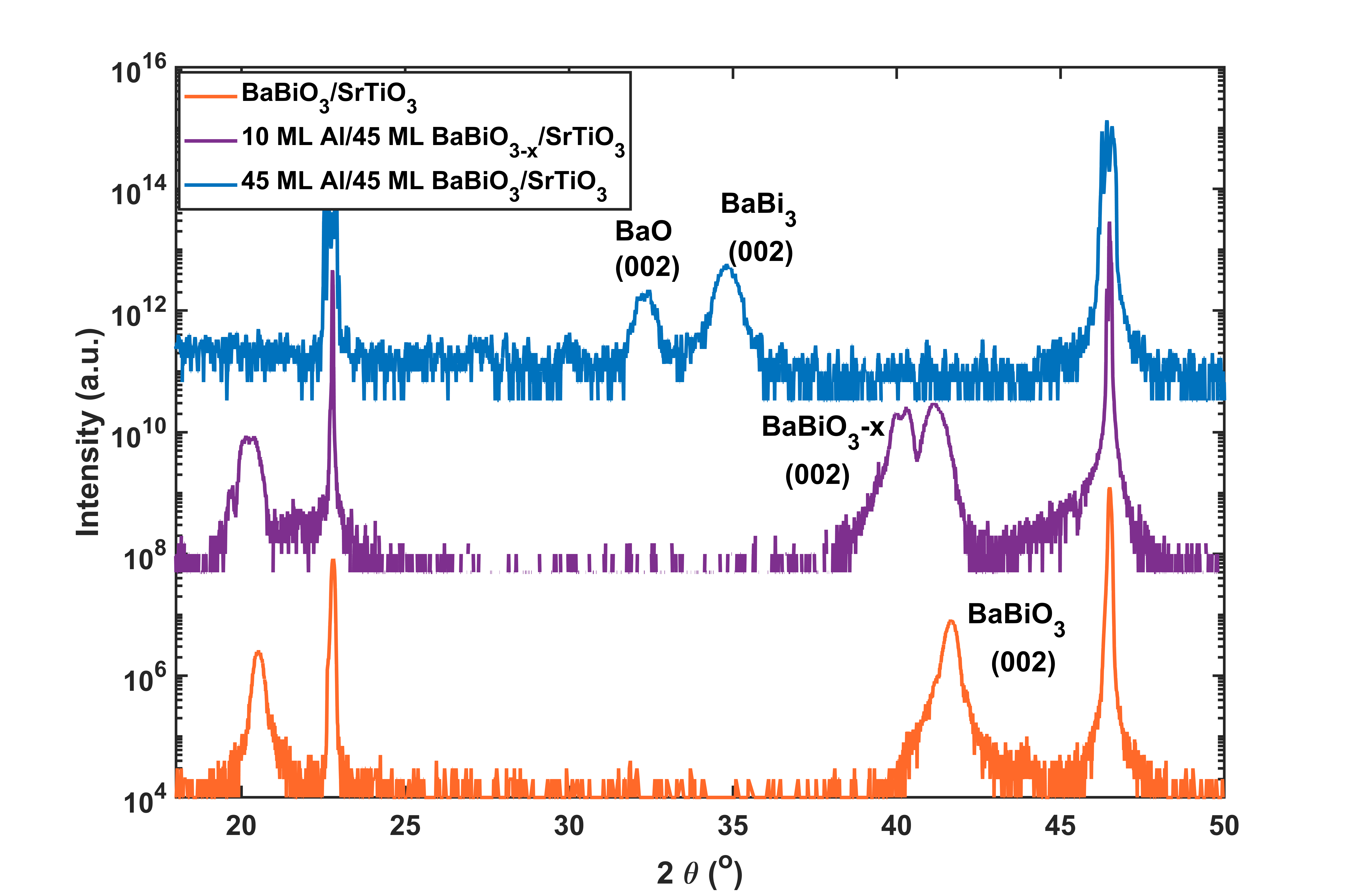} 

	\caption{\textbf{Effect of Al cap thickness on reduction of BaBiO$_3$ films grown on (001)-oriented SrTiO$_3$. The perovskite BaBiO$_3$ phase is confirmed for uncapped films. Partial reduction of the BaBiO$_3$ occurs when the Al overlay thickness is less than the BaBiO$_3$ thickness. Decomposition of the BaBiO$_3$ to BaO and superconducting BaBi$_3$ is observed for 1:1 ratios of the Al and BaBiO$_3$. }}
	\label{fig:Althickness} 
\end{figure*}

\begin{figure*} 
	\centering
	\includegraphics[width=1.0\textwidth]{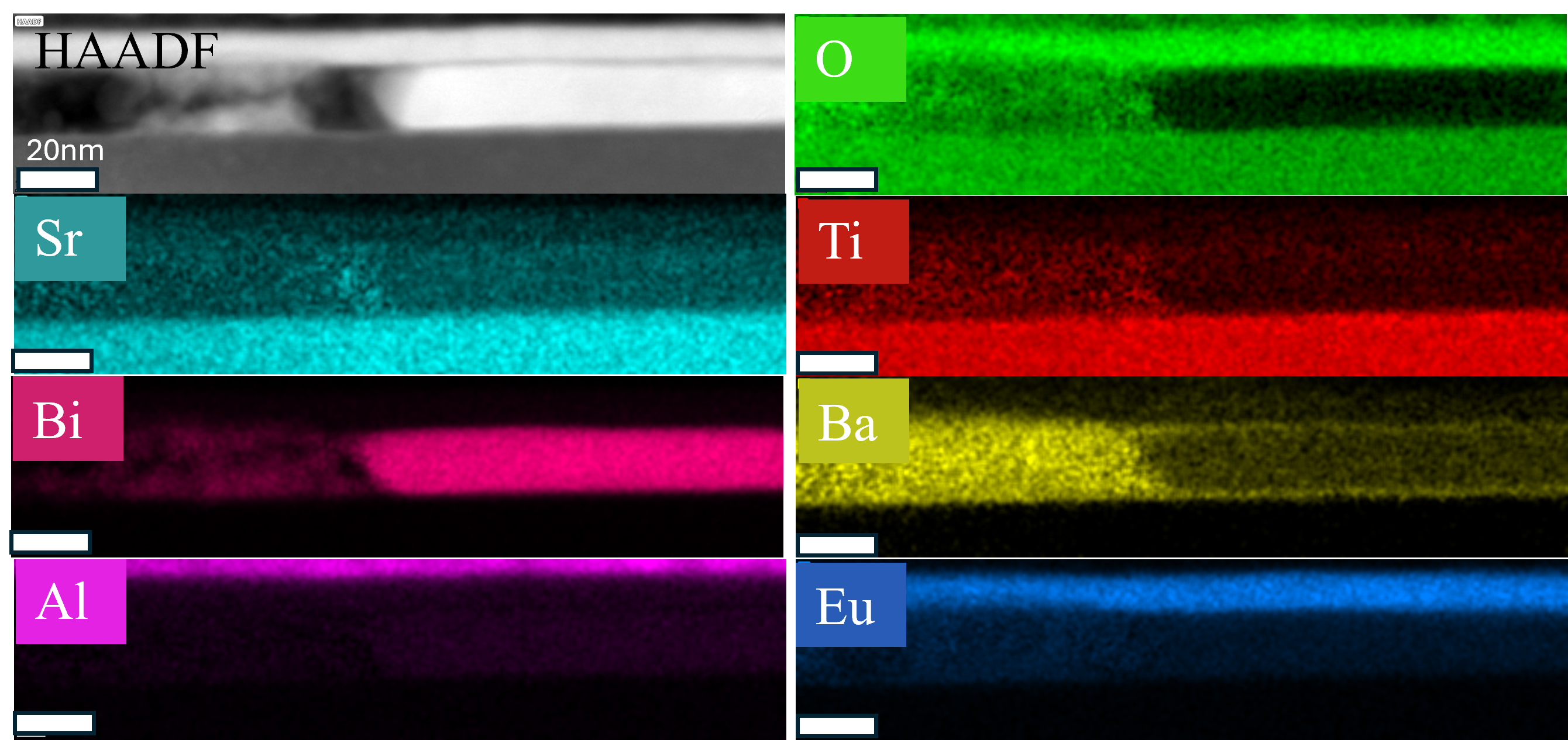} 

	\caption{\textbf{HAADF and EDX maps showing regions of film with lateral segregation of BaO and BaBi$_3$. The scale bar corresponds to 20 nm.}}
	\label{fig:sup_example} 
\end{figure*}

\begin{figure*} 
	\centering
	\includegraphics[width=1.0\textwidth]{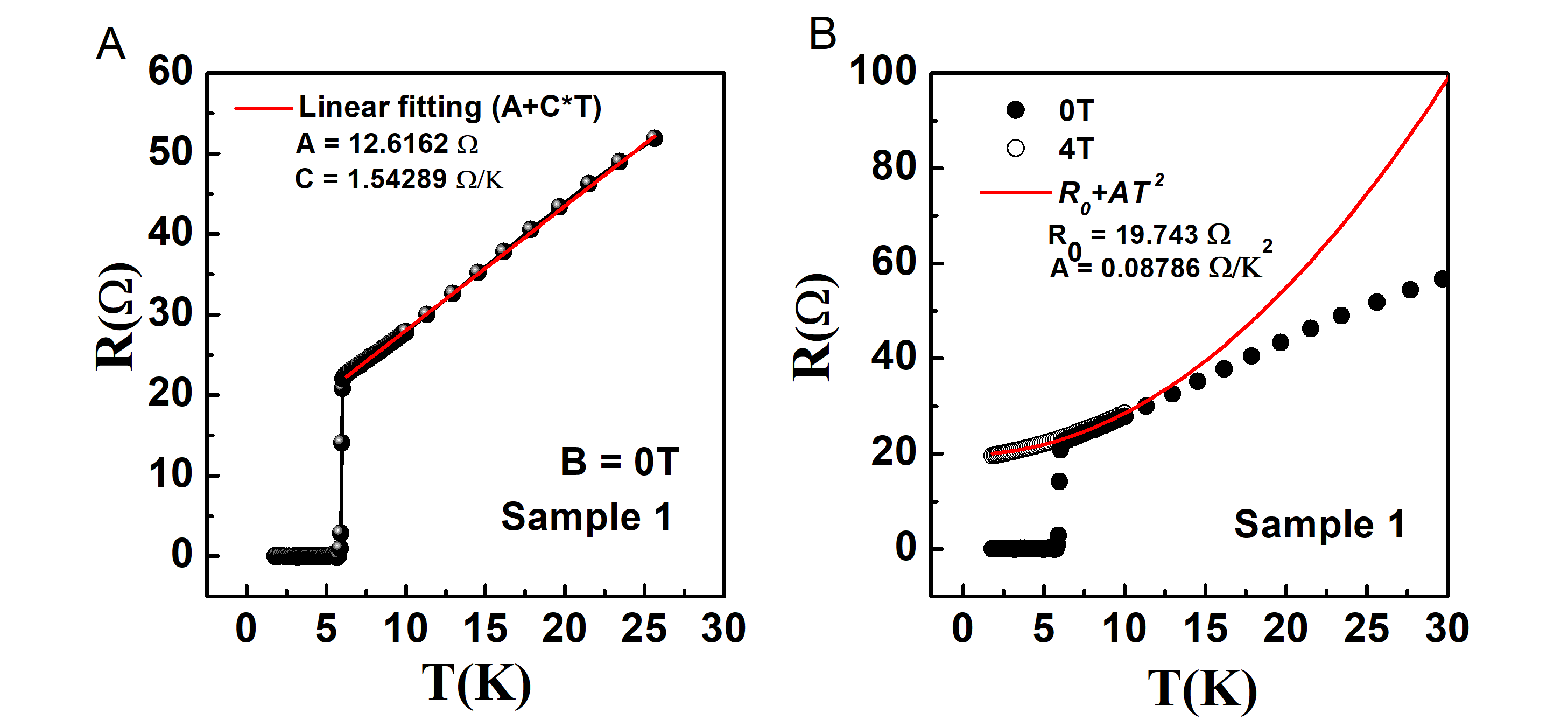} 

	\caption{\textbf{(A) Resistance vs. temperature for Sample 1, with the red line indicating a linear fit to the normal state resistance characteristic of strange metal behavior. (B) Resistance as a function of temperature at B = 0 and 4T. The red line represents the fit using R = R$_{0}$ + AT$^{2}$. }}
	\label{fig:Res} 
\end{figure*}

\begin{figure*} 
	\centering
	\includegraphics[width=1.0\textwidth]{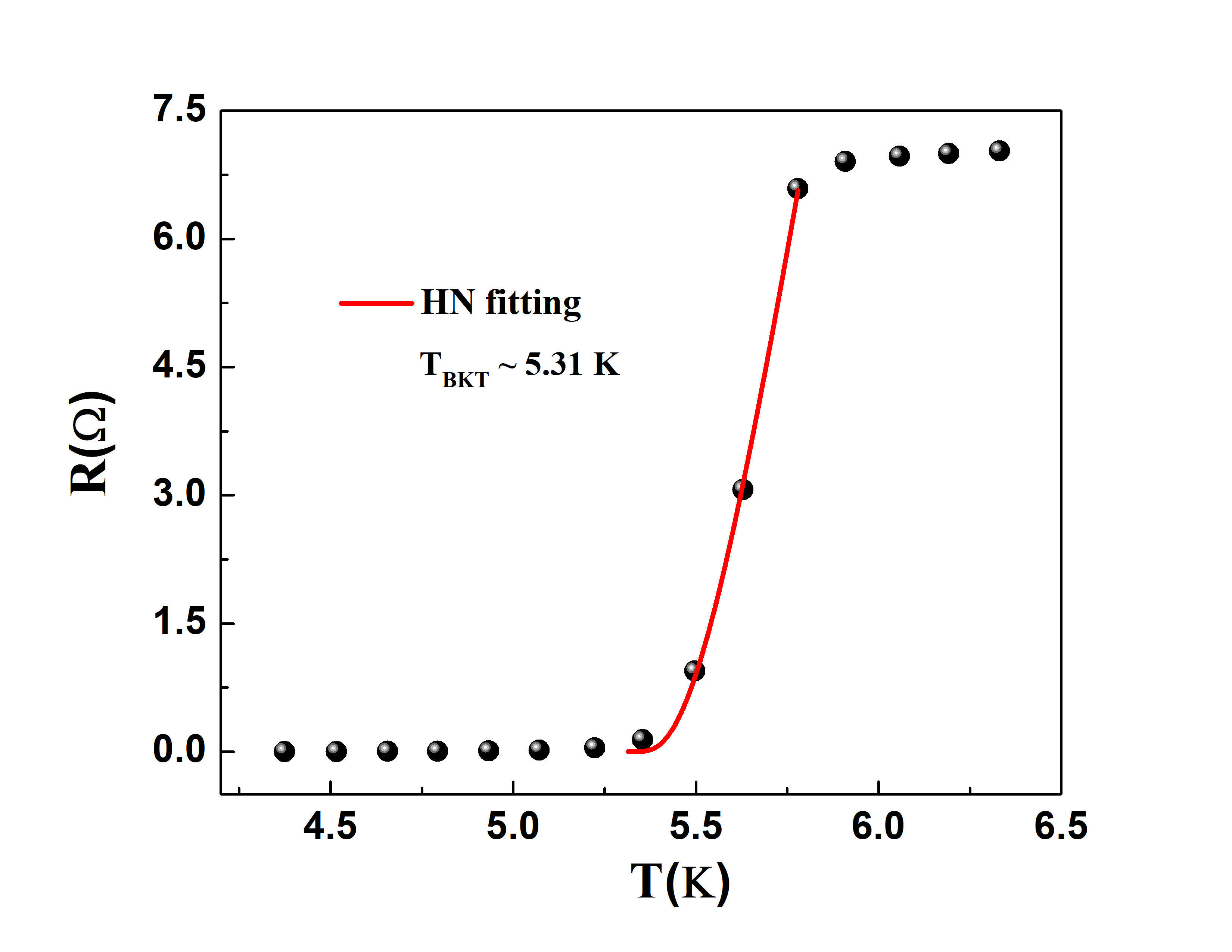} 

	\caption{\textbf{Resistance as a function of temperature for Sample 2. The red solid line represents a fit based on the Halperin-Nelson formula.}}
	\label{fig:BKT:sup} 
\end{figure*}

\begin{figure*} 
	\centering
	\includegraphics[width=1.0\textwidth]{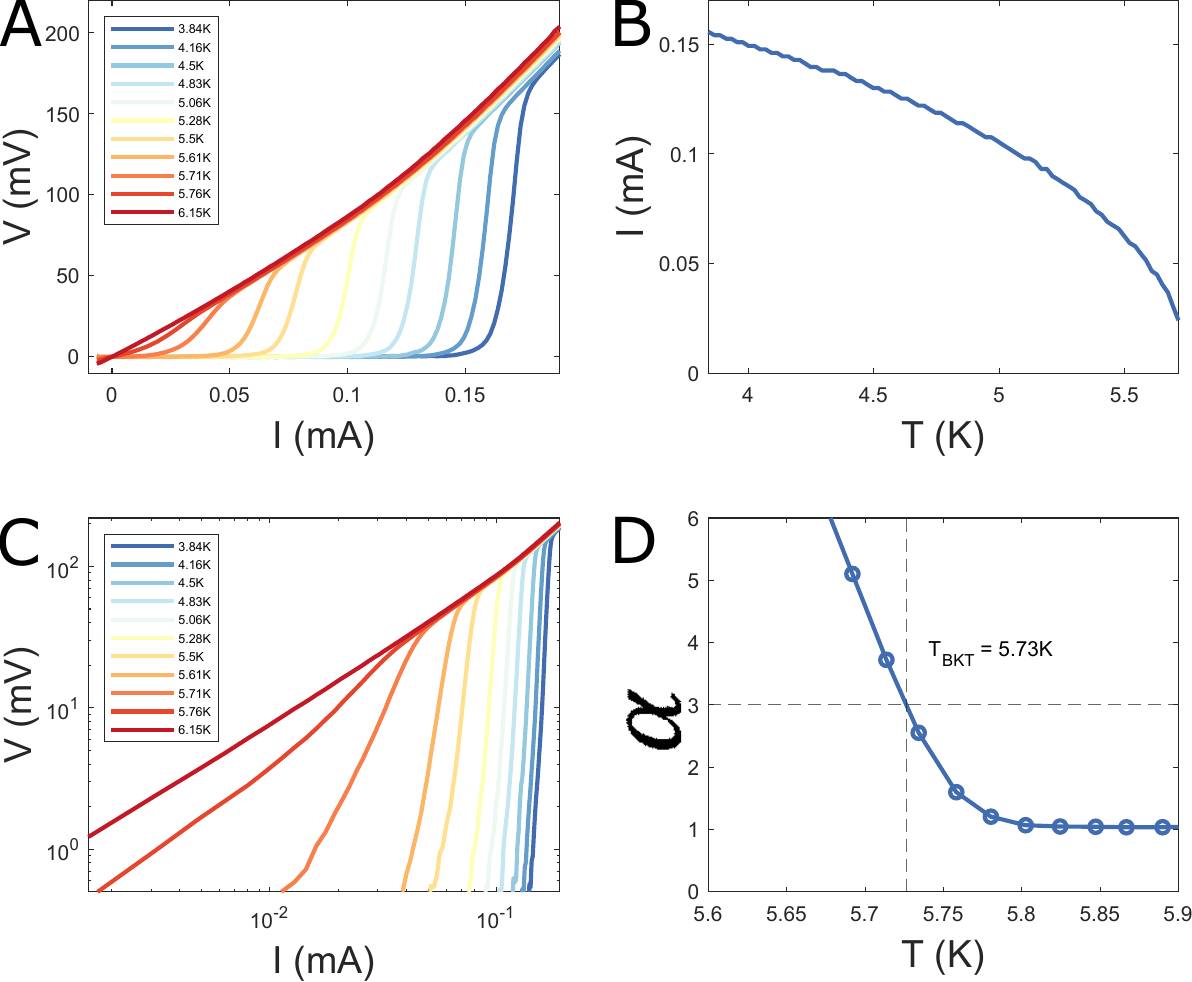} 

	\caption{\textbf{(A) I-V measurements of a nominal Eu/p-BBO (Sample 3) film grown on (00)-oriented SrTiO$_3$. (B) $I_c$ determined from the I-V curves as a function of temperature. $I_c$ is defined as the maximum derivative ($dV/dI$) at a given temperature. (C) I-V curves plotted on a logarithmic scale. (D) Extracted slope ($\alpha$) from fitting the logarithm of the I-V curves. A BKT transition temperature of 5.73 K is extracted.}}
	\label{fig:BKT_JM:sup} 
\end{figure*}

\end{document}